\documentclass[a4paper]{article}

\usepackage{mathtools}
\usepackage{graphicx}
\usepackage{physics}
\usepackage{color}
\usepackage{subcaption}
\usepackage{amssymb}
\usepackage{relsize}
\usepackage{hyperref}
\usepackage[colorinlistoftodos]{todonotes}
\usepackage{grffile}
\usepackage{cuted}
\usepackage{rsfso}
\usepackage[bottom]{footmisc}
\usepackage{placeins}
\usepackage{authblk}

\usepackage[a4paper,top=3cm,bottom=2cm,left=3cm,right=3cm,marginparwidth=1.75cm]{geometry}

\newcommand\myeq{\mathrel{\stackrel{\makebox[0pt]{\mbox{\normalfont\tiny\sffamily def}}}{=}}}

\let\baraccent=\= 

\newcommand{\bmx}{\left( \begin{matrix}}
\newcommand{\emx}{\end{matrix} \right)}
\newcommand{\bsm}{\begin{psmallmatrix}}
\newcommand{\esm}{\end{psmallmatrix}}

\begin{document}

\title{Nonparaxial Cartesian and azimuthally symmetric waves with concentrated wavevector and frequency spectra}

\author[1]{Mateus Corato-Zanarella\footnote{Corresponding author: mateuscorato@gmail.com}}
\author[2]{Henrique Corato-Zanarella}
\author[1]{Michel Zamboni-Rached}
\affil[1]{School of Electrical and Computer Engineering, University of Campinas, Campinas, SP, Brazil}
\affil[2]{Independent collaborator}

\date{}

\maketitle

\begin{abstract}

\par In this paper, we develop a theoretical analysis to efficiently handle superpositions of waves with concentrated wavevector and frequency spectra, allowing an easy analytical description of fields with interesting transverse profiles. First, we analyze an extension of the paraxial formalism that is more suitable for superposing these types of waves, as it does not rely on the use of coordinate rotations combined with paraxial assumptions. Second, and most importantly, we leverage the obtained results to describe azimuthally symmetric waves composed of superpositions of zero-order Bessel beams with close cone angles that can be as large as desired, unlike in the paraxial formalism. Throughout the paper, examples are presented, such as Airy beams with enhanced curvatures, nonparaxial Bessel-Gauss beams and Circular Parabolic-Gaussian beams (which are based on the Cartesian Parabolic-Gaussian beams), and experimental data illustrates interesting transverse patterns achieved by superpositions of beams propagating in different directions.

\end{abstract}

\section{Introduction}

\par Appropriate superpositions of waves propagating in different directions may create interesting and useful interference patterns in the transverse plane that keep their shape up to a certain propagation distance, which is the range within which the waves interfere significantly. Consequently, it is higher when the spot sizes of the superposed waves are bigger, which implies that their wavevector spectra are concentrated around their propagation directions. Since these can make large angles with respect to the longitudinal axis, the paraxial wave equation may not describe the propagation of the superposition properly. 

\par In principle, each wave could be analyzed separately using the paraxial formalism in a system of rotated coordinates whose z-axis coincide with its propagating direction and the application of the inverse rotation would express the wave as a function of the desired non-rotated coordinates. However, this procedure is not straightforward and results in complicated expressions, as the rotations mix the transverse and longitudinal non-rotated coordinates. 

\par To allow a more suitable and analytically simple way of handling superpositions of waves with concentrated spectra, we analyze and apply an extension of the paraxial formalism that deals with combinations of plane waves whose wavevectors are close to any desired direction without relying on coordinate rotations. The resulting expressions do not mix transverse and longitudinal coordinates while still incorporate the inclination of the plane waves very clearly through parameters. Moreover, if desired, the method can be straightforwardly extended to include higher-order corrections, so it is not limited to the accuracy of the paraxial approximation.

\par As the main contribution of this work, we then leverage these results to develop a novel analytical treatment for azimuthally symmetric waves with concentrated spectra, that is, waves composed of superpositions of zero-order Bessel beams with close cone angles. However, in contrast to the paraxial assumption of small angles with respect to the longitudinal axis, this formalism encompasses cone angles as large as desired and provides simple expressions for this class of nonparaxial waves. In particular, similarities with the unidimensional\footnote{We will say that a wave is n-dimensional when it is a function of n transverse spatial coordinates, in addition to the longitudinal coordinate $z$ and the time $t$.} Cartesian waves here analyzed allow us to build azimuthally symmetric versions of them, which can be used to derive new types of waves. As examples, we provide an analytical expression for nonparaxial Bessel-Gauss beams and for azimuthally symmetric versions of Parabolic-Gaussian beams (which we refer to as Circular Parabolic-Gaussian beams).

\par Throughout this work, we also present experimental demonstrations and comparisons to numerically-calculated Rayleigh-Sommerfeld diffraction integrals to corroborate the validity and accuracy of the theoretical results.

\section{Integral formulation and wave equation}
\label{integral_formulation_and_PDE}

\subsection{Bidimensional case}
\label{bidimensional}

\par Let $\Psi(x,y,z,t)$ be an arbitrary solution of the homogeneous scalar wave equation $(\nabla^2-\partial_{ct}^2)\Psi=0$. It can be represented by a superposition of plane waves with different wavevectors $\vec{k}=k_x\hat{x}+k_y\hat{y}+k_z\hat{z}$ and angular frequencies $\omega$, given that $|\vec{k}|=k(\omega)=\omega/c$, where $c$ is the speed of light in the medium. If we consider only waves that propagate in the $+\hat{z}$ direction, $k_z=\sqrt{k^2(\omega)-(k_x^2+k_y^2)}$ and $\Psi(x,y,z,t)$ can be written as
\begin{equation}
\Psi(x,y,z,t)=\int_{-\infty}^{\infty} \text{d}\omega \int_{-\infty}^{\infty} \text{d}k_x \int_{-\infty}^{\infty} \text{d}k_y \, \tilde{S}(k_x,k_y,\omega) e^{-i\omega t} e^{ik_x x} e^{ik_y y} e^{i \sqrt{k^2(\omega)-(k_x^2+k_y^2)}z}
\label{exact_superposition}
\end{equation}

\noindent where $\tilde{S}(k_x,k_y,\omega)$ is the spectrum, which determines the amplitude and phase of each plane wave. The limits of the integrals in Eq. \eqref{exact_superposition} allow imaginary values of $k_z$, accounting for the presence of evanescent waves. If only propagating waves are desired, the integration should be performed in the region $k_x^2+k_y^2\leq\omega^2/c^2$.

\par If the spectrum $\tilde{S}(k_x,k_y,\omega)$ is concentrated around $(k_x,k_y,\omega)=(k_{x_0},k_{y_0},\omega_0)$, the plane waves have wavevectors that vary only slightly around $\vec{k}_0=k_{x_0}\hat{x}+k_{y_0}\hat{y}+\sqrt{k_0^2-(k_{x_0}^2+k_{y_0}^2)}\hat{z}$, where $k_0\myeq k(\omega_0)=\omega_0/c$. Mathematically, this means that $|\tilde{S}(k_x,k_y,\omega)|$ is significant only in the region in which $|k_x^{\prime}/k_{x_0}| \ll 1$, $|k_y^{\prime}/k_{y_0}| \ll 1$ and $|u/\omega_0| \ll 1$, with $k_x^{\prime}\myeq k_x-k_{x_0}$, $k_y^{\prime}\myeq k_y-k_{y_0}$ and $u\myeq \omega-\omega_0$ being defined as the deviation variables \footnote{Additionally, we may assume that $k_{x_0}<k_0$, $k_{y_0}<k_0$ and $k_{x_0}^2+k_{y_0}^2<k_0^2$, so that all the waves are propagating. In this case, even though we keep the limits of the integrals in $k_x$ and $k_y$ from $-\infty$ to $\infty$, the concentration of the spectrum implies that there are no significant evanescent waves in the superposition.}. In this region, $k_z$ can be approximated to second order in $k_x^\prime$ and $k_y^\prime$ using a Taylor series expansion\footnote{If desired, higher-order corrections can be easily included to improve accuracy, even though the second-order approximation is usually enough.}:
\begin{align}
\nonumber k_z  =& \sqrt{k^2(\omega)-k_x^2-k_y^2} \approx \\
\nonumber &\sqrt{k^2(\omega)-k_{\mathsmaller{\perp}_0}^2} -\frac{k_{x_0}}{\sqrt{k^2(\omega)-k_{\mathsmaller{\perp}_0}^2}}k^{\prime}_x -\frac{k_{y_0}}{\sqrt{k^2(\omega)-k_{\mathsmaller{\perp}_0}^2}}k^{\prime}_y \\
&-\frac{k^2(\omega)}{2\left[k^2(\omega)-k_{\mathsmaller{\perp}_0}^2\right]^\frac{3}{2}} {k^{\prime}_x}^2 - \frac{k^2(\omega)}{2\left[k^2(\omega)-k_{\mathsmaller{\perp}_0}^2\right]^\frac{3}{2}} {k^{\prime}_y}^2 -\frac{{k_{x_0}}{k_{y_0}}}{\left[k^2(\omega)-k_{\mathsmaller{\perp}_0}^2\right]^\frac{3}{2}} {k^{\prime}_x}{k^{\prime}_y}
\label{kz_approx}
\end{align}

\noindent where $k_{\mathsmaller{\perp}_0}^2\myeq k_{y_0}^2+k_{x_0}^2$. For the variation in $\omega$, we perform a first order Taylor series approximation only in the most significant term, which is the first one on the right-hand side of Eq. \eqref{kz_approx}, resulting in
\begin{equation}
\sqrt{k^2(\omega)-k_{\mathsmaller{\perp_0}}^2} \approx \sqrt{k_0^2-k_{\mathsmaller{\perp}_0}^2} + \frac{u}{c}\frac{k_0}{\sqrt{k_0^2-k_{\mathsmaller{\perp}_0}^2}}
\end{equation}

\par In all the other terms, we take $k(\omega) \approx k_0$. Assigning names to the constants of Eq. \eqref{exact_superposition} and Eq. \eqref{kz_approx}:
\begin{subequations}
\begin{gather}
a_x\myeq\frac{k_{x_0}}{\sqrt{k_0^2-k_{\mathsmaller{\perp}_0}^2}} \,\text{,} \quad a_y\myeq\frac{k_{y_0}}{\sqrt{k_0^2-k_{\mathsmaller{\perp_0}}^2}} \\
b \myeq \frac{1}{2}\frac{k_0^2}{\left(k_0^2-k_{\mathsmaller{\perp_0}}^2\right)^{\frac{3}{2}}} \\
d \myeq \frac{k_{x_0}k_{y_0}}{\left(k_0^2-k_{\mathsmaller{\perp_0}}^2\right)^{\frac{3}{2}}}\\
e \myeq \frac{k_0}{\sqrt{k_0^2-k_{\mathsmaller{\perp_0}}^2}}
\end{gather}
\end{subequations}

\noindent and performing the integrals of Eq. \eqref{exact_superposition} on the deviation variables, we get
\begin{equation}
\Psi(x,y,z,t) \approx  e^{-i\omega_0t}e^{ik_{x_0}x}e^{ik_{y_0}y}e^{i\sqrt{k_0^2-k_{\mathsmaller{\perp_0}}^2}z}A(x,y,z,t)
\label{Psi_2D}
\end{equation}
\begin{equation}
A(x,y,z,t) \myeq  \int_{-\infty}^{+\infty}\text{d}u\int_{-\infty}^{+\infty}\text{d}k^\prime_x\int_{-\infty}^{+\infty}\text{d}k^{\prime}_y \, S(k^{\prime}_x,k^{\prime}_y,u) e^{-iut}e^{ik^{\prime}_x x}e^{ik^{\prime}_y y}e^{i k^{\prime}_z z}
\label{A_2D}
\end{equation}

\noindent where $S(k^{\prime}_x,k^{\prime}_y,u)\myeq \tilde{S}(k^{\prime}_x+k_{x_0}, k^{\prime}_y+k_{y_0}, u+\omega_0)$ and
\begin{equation}
k^{\prime}_z=-a_x k^{\prime}_x -a_y k^{\prime}_y -d k^{\prime}_x k^{\prime}_y -b(k_x^{\prime\, 2}+k_y^{\prime\, 2}) +\frac{e}{c}u
\label{kz_prime_approx}
\end{equation}

\noindent is the deviation of $k_z$ with respect to $k_{z_0}\myeq \sqrt{k_0^2-k_{\mathsmaller{\perp_0}}^2}$.

\par Eq. \eqref{Psi_2D} shows that $\Psi(x,y,z,t)$ is approximately a plane wave with wavevector $\vec{k}_0$ and frequency $\omega_0$ modulated by an envelope $A(x,y,z,t)$ which, according to Eq. \eqref{A_2D}, is composed of a superposition of plane waves with wavevectors $\vec{k}^\prime=k^{\prime}_x\hat{x}+k^{\prime}_y\hat{y}+k^{\prime}_z\hat{z}$ and angular frequencies $u$. Since $|k^{\prime}_x|$, $|k^{\prime}_y|$, $|k^{\prime}_z|$ and $|u|$ are small compared to $|k_{x_0}|$, $|k_{y_0}|$, $|k_{z_0}|$ and $\omega_0$, respectively, $A(x,y,z,t)$ has slow variations in $x$, $y$, $z$ and $t$. Mathematically, this means
\begin{subequations}
\begin{equation}
\left|\frac{\partial A}{\partial x_i}\right| \ll \left|k_{x_{i_0}}A\right|\,\text{,} \quad \left|\frac{\partial^2 A}{\partial x_i^2}\right| \ll \left|k_{x_{i_0}}\frac{\partial A}{\partial x_i}\right|
\label{inequalities_space}
\end{equation}
\begin{equation}
\left|\frac{\partial A}{\partial t}\right| \ll |\omega_0 A|\,\text{,} \quad \left|\frac{\partial^2 A}{\partial t^2}\right| \ll \left|\omega_0\frac{\partial A}{\partial t}\right|
\label{inequalities_time}
\end{equation}
\label{inequalities}
\end{subequations}
\vspace{1mm}
\noindent where $x_i$ stands for any spatial coordinate.

\par The differential equation satisfied by $A(x,y,z,t)$ is uniquely defined by the relation among the frequency and wavevector components of its composing plane waves, given by Eq. \eqref{kz_prime_approx}. Each differentiation of $A(x,y,z,t)$ with respect to a variable implies a multiplication of its spectrum by a factor proportional to the corresponding reciprocal variable, resulting in the correspondences:
\begin{equation}
\frac{1}{i}\frac{\partial}{\partial x_i} \longleftrightarrow k^\prime_{x_i} \, \text{,}\quad -\frac{\partial^2}{\partial x_i^2} \longleftrightarrow k_{x_i}^{\prime\,2} \, \text{,} \quad -\frac{1}{i}\frac{\partial}{\partial t} \longleftrightarrow u
\end{equation}

\noindent where $x_i$ stands for any spatial coordinate. Therefore, Eq. \eqref{kz_prime_approx} is equivalent to
\begin{equation}
\frac{\partial A}{\partial z} = -\frac{e}{c}\frac{\partial A}{\partial t} - \vec{a}\cdot\vec{\nabla}_\perp A + ib\nabla_\perp^2 A +id\frac{\partial^2 A}{\partial x \partial y}
\label{A_eq}
\end{equation}

\noindent where $\vec{a}\myeq a_x\hat{x}+a_y\hat{y}$ and $\vec{\nabla}_\perp \myeq \hat{x}\partial_x + \hat{y}\partial_y$. The coefficient of the mixed derivative in Eq. \eqref{A_eq} implies that the equation has separable solutions of the form $\Psi(x,y,z,t)=f(x,z)g(y,z)h(t)$ only when $d=0$, that is, when at least one of $k_{x_0}$ and $k_{y_0}$ is zero. If they are both zero, $\vec{a}=\vec{0}$, $b=(2k_0)^{-1}$, $d=0$ and $e=1$ and Eq. \eqref{A_eq} reduces to the well-known paraxial wave equation $\frac{\partial A}{\partial z}=-\frac{1}{c}\frac{\partial A}{\partial t}+\frac{i}{2k_0}\nabla_\perp^2 A$, since in this case the superposed plane waves have wavevectors almost parallel to the z-direction.

\subsection{Unidimensional case}

\par If $\Psi$ is a function of only one of the transverse coordinates, say $\Psi(x,z,t)$, the same kind of analysis provides results similar to those of sec. \ref{bidimensional}:
\begin{gather}
\Psi(x,z,t)  \approx e^{-i\omega_0t}e^{ik_{x_0}x}e^{i\sqrt{k_0^2-k_{\mathsmaller{\perp}_0}^2}z}A(x,z,t) \label{Psi_1D} \\
A(x,z,t)  = \int_{-\infty}^{+\infty}\text{d}u\int_{-\infty}^{+\infty}\text{d}k^\prime_x \, S(k^{\prime}_x,u) e^{-iut}e^{ik^{\prime}_x x}e^{i k^{\prime}_z z} \label{A_1D} \\
k^{\prime}_z  = -a_x k^{\prime}_x -bk_x^{\prime\, 2} +\frac{e}{c}u \\
\frac{\partial A}{\partial z} = -\frac{e}{c}\frac{\partial A}{\partial t} - a_x \frac{\partial A}{\partial x}+ ib\frac{\partial^2 A}{\partial x^2} \label{A_eq_1D}
\end{gather}

\noindent with $k_{\mathsmaller{\perp}_0}=k_{x_0}$. In this case, it is possible to write Eq. \eqref{A_eq_1D} in a dimensionless form by defining new constants and dimensionless variables:
\begin{gather}
\gamma \myeq \frac{k_{x_0}}{k_0} \,\text{,} \quad \alpha \myeq k_{x_0}x_0 \label{constantes_caso_adimensional}\\
s \myeq \frac{x}{x_0} \,\text{,} \quad \xi \myeq \frac{z}{k_0x_0^2}\frac{1}{(1-\gamma^2)^\frac{3}{2}} \,\text{,} \quad \tau \myeq \frac{c}{k_0x_0^2}t \label{variaveis_adimensionais}
\end{gather}

\noindent where $x_0$ is a constant. Then, the differential equation for $A(s,\xi,\tau)$ becomes
\begin{equation}
\frac{\partial^2 A}{\partial s^2} + 2i\frac{\partial A}{\partial \xi} + 2i\alpha(1-\gamma^2)\frac{\partial A}{\partial s} + 2i(1-\gamma^2)\frac{\partial A}{\partial \tau} = 0
\end{equation}

\par When $k_{x_0}=0$ it correctly reduces to the dimensionless unidimensional paraxial wave equation $\frac{\partial^2 A}{\partial s^2} + 2i\frac{\partial A}{\partial \xi} + 2i\frac{\partial A}{\partial \tau} = 0$.

\par Using these constants and dimensionless variables and defining the dimensionless wavevector $\tilde{k}_{x}^\prime\myeq k_x^\prime x_0$, the wave of Eq. \eqref{Psi_1D} can be written as
\begin{align}
\nonumber \Psi(s,\xi,\tau) \approx &e^{-i\frac{\alpha^2}{\gamma^2}\tau} e^{i\alpha s} e^{i\frac{\alpha^2}{\gamma^2}(1-\gamma^2)^2\xi} \int_{-\infty}^{+\infty} \text{d}u \int_{-\infty}^{+\infty} \text{d}\tilde{k}_x^\prime \, \mathcal{S}(\tilde{k}^{\mathsmaller\prime}_x,u)  \\
		& \times e^{-i\frac{\alpha^2}{\omega_0 \gamma^2}u\tau} e^{i\tilde{k}_x^\prime s} e^{-i\left[\alpha(1-\gamma^2) \tilde{k}^{\prime}_x + \frac{1}{2}\tilde{k}_x^{\prime \, 2}-\frac{\alpha^2}{\gamma^2}(1-\gamma^2)\frac{u}{\omega_0}\right]\xi} 
\end{align}

\noindent where $\mathcal{S}(\tilde{k}^{\mathsmaller\prime}_x,u) \myeq S(\tilde{k}^{\mathsmaller\prime}_x/x_0,u)/x_0$ is a concentrated spectrum around $(\tilde{k}^{\mathsmaller\prime}_x,u)=(0,0)$.

\subsection{Reduction to the paraxial wave equation}
\label{seção de redução à equação de onda}

\par By means of some changes of variables, it is possible to reduce the general Eq. \eqref{A_eq} to the paraxial wave equation, so we can leverage all we know about this equation to Eq. \eqref{A_eq}. The first order derivatives in $x$, $y$ and $t$ can be eliminated if we introduce the shifted variables $\overline{x}$, $\overline{y}$ and $\overline{t}$ according to
\begin{subequations}
\begin{align}
\overline{x} &= x-a_x z\\
\overline{y} &= y-a_yz \\
\overline{t} &= t-\frac{e}{c}z \\
\overline{z} &= z
\end{align}
\end{subequations}

\par Applying the chain rule and defining $\vec{\nabla}^\prime_\perp \myeq \hat{x}{\partial}_{\overline{x}} + \hat{y}{\partial}_{\overline{y}}$, Eq. \eqref{A_eq} can be rewritten as
\begin{equation}
\frac{\partial A}{\partial \overline{z}}=i\mathcal{L}[A]\,\text{,} \quad \mathcal{L}\myeq b\nabla^{\prime\,2}_\perp + d\frac{\partial^2}{\partial \overline{x} \partial \overline{y}}
\label{A_reduc_1}
\end{equation}

\par Note that the shifts in the transverse coordinates express the inclination of $\vec{k}_0$ with respect to the z-direction and the shift in the temporal variable represents the wave's travel time to arrive at the longitudinal position $z$. Eq. \eqref{A_reduc_1} shows that any solution of Eq. \eqref{A_eq} must be of the form $A(x-a_x z,y-a_y z, z, t-ez/c)$.

\par With an additional change of variables of the type
\begin{equation}
\zeta=a_1 \overline{x} + a_2 \overline{y} \,\text{,} \quad \eta = a_3 \overline{x} + a_4 \overline{y}
\end{equation}

\noindent it is possible to eliminate the mixed derivative of the operator $\mathcal{L}$. Two alternatives that reduce it to $\mathcal{L}=b\left[\frac{\partial^2}{\partial \zeta^2}+\frac{\partial^2}{\partial \eta^2}\right]$ are
\begin{subequations}
\begin{align}
\zeta &=\frac{1}{\sqrt{4b^2-d^2}} (2b\overline{x} - d\overline{y}) \\
\eta &=\overline{y}
\end{align}
\label{transformation_1}
\end{subequations}

\noindent and
\begin{subequations}
\begin{align}
\zeta=\frac{1}{\sqrt{2(4b^2-d^2)}} [2b\overline{x} - (d+\sqrt{4b^2-d^2}) \overline{y}] \\
 \eta= \frac{1}{\sqrt{2(4b^2-d^2)}}[2b\overline{x} - (d-\sqrt{4b^2-d^2})\overline{y}]
\end{align}
\label{transformation_2}
\end{subequations}

\noindent so that Eq. \eqref{A_reduc_1} becomes
\begin{equation}
\frac{\partial A}{\partial \overline{z}}=ib\left[\frac{\partial^2 A}{\partial \zeta^2}+\frac{\partial^2 A}{\partial \eta^2}\right]
\label{paraxial_xi_eta}
\end{equation}

\noindent which is like a paraxial wave equation in the variables $\zeta$ and $\eta$. The factor $b$ was not changed to $(2k_0)^{-1}$ because, as already pointed out in sec. \ref{bidimensional}, $b=(2k_0)^{-1}$ if $k_{x_0}=k_{y_0}=0$. However, if desired, an additional scaling in both $\zeta$ and $\eta$ can reduce Eq. \eqref{paraxial_xi_eta} precisely to the paraxial wave equation in the new scaled variables, which are $\overline{\zeta}=\zeta/\sqrt{2bk_0}$ and $\overline{\eta}=\eta/\sqrt{2bk_0}$. In the unidimensional case, Eq. \eqref{A_eq_1D} may be reduced to $\partial_{\overline{z}} A=ib\partial_{\overline{x}\overline{x}}A$ and $A$ also obeys the paraxial wave equation in the scaled variable $\chi=\overline{x}/\sqrt{2bk_0}$.

\par Therefore, any solution of the paraxial wave equation can be extended to 
the case being considered, which is natural, since this approach should be analogous to a rotation of coordinates. Note, however, that the former is simpler: while the rotation expresses the $z$ coordinate as a combination of both transverse and longitudinal rotated coordinates, the presented method keep the transverse and longitudinal coordinates separated.

\par The choice of Eq. \eqref{transformation_1} for $\zeta$ and $\eta$, although asymmetrical in $\overline{x}$ and $\overline{y}$, reduces to $\zeta=\overline{x}$ and $\eta=\overline{y}$ when the ``coupling factor'' $d$ is zero, showing that the lack of the mixed derivative makes it possible to find a separable solution for $A$ in the variables $\overline{x}$ and $\overline{y}$. The choice of Eq. \eqref{transformation_2}, on the other hand, has a better symmetry in $\overline{x}$ and $\overline{y}$, but reduces to $\zeta=(\overline{x}-\overline{y})/\sqrt{2}$ and $\eta=(\overline{x}+\overline{y})/\sqrt{2}$ when $d=0$.

\par Also, note that if a beam with envelope $A(x,y,z)$ satisfies Eq. \eqref{A_eq}, the pulse with envelope $\overline{A}(x,y,z,t)=A(x,y,z)B(t-ez/c)$ is also a solution of that equation for an arbitrary differentiable function $B(t-ez/c)$.

\section{Diffraction integral}

\par For waves with concentrated spectrum around $(k_x,k_y,\omega)=(k_{x_0}, k_{y_0}, \omega_0)$ it is possible to derive a generalization of the Fresnel diffraction integral.

\subsection{Bidimensional case}

\par From Eq. \eqref{A_2D}, it is evident that $A(x,y,0,t)$ is a Fourier transform of $S(k^\prime_x,k^\prime_y,u)$. Therefore, by inverting the transform, we have
\begin{equation}
S(k^{\prime}_x,k^{\prime}_y,u) =\frac{1}{(2\pi)^3}\int_{-\infty}^{+\infty}\text{d}x^{\prime}\int_{-\infty}^{+\infty}\text{d}y^{\prime}\int_{-\infty}^{+\infty}\text{d}t^{\prime} A(x^{\prime},y^{\prime},0,t^{\prime})e^{-ik^{\prime}_x x^\prime}e^{-ik^{\prime}_y y^\prime}e^{iut^{\prime}}
\label{S_A0}
\end{equation}

\par Substituting Eq. \eqref{S_A0} in Eq. \eqref{A_2D}, the integrals in $k^{\prime}_x$, $k^{\prime}_y$ and $u$ can be solved by completing squares and using the known results $\int_{-\infty}^{+\infty}e^{-ix^2} \text{d}x= (1-i)\sqrt\frac{\pi}{2}$ and $\int_{-\infty}^{+\infty}e^{iux}\text{d}u = 2\pi\delta(x)$, where $\delta(\cdot)$ is the Dirac delta function. This leads to the diffraction integral:
\begin{align}
\nonumber &A(x,y,z,t) = \frac{1}{4{\pi}ibz\sqrt{1-\frac{d^2}{4b^2}}}\int_{-\infty}^{+\infty}\text{d}x^{\prime}\int_{-\infty}^{+\infty}\text{d}y^{\prime}A\left(x^{\prime},y^{\prime},0,t-\frac{e}{c}z\right)\\
& \times \exp\left[{\frac{i}{4bz(1-\frac{d^2}{4b^2})}\left((x-x^{\prime}-a_xz)^2 + (y-y^{\prime}-a_yz)^2 - \frac{d}{b}(x-x^{\prime}-a_xz)(y-y^{\prime}-a_yz)\right)}\right]
\label{diff_2D}
\end{align}

\par For beams, that is, when $A=A(x,y,z)$ is not a function of time, Eq. \eqref{diff_2D} is still valid if we remove the $t$ dependencies. As it should be, for $k_{x_0}=k_{y_0}=0$ it reduces to the Fresnel diffraction integral. An alternative procedure to derive Eq. \eqref{diff_2D} is to write the Fresnel diffraction integral for the transformed variables $\overline{\zeta}$ and $\overline{\eta}$ and arrange the terms in $x-a_x z$ and $y-a_y z$.

\subsection{Unidimensional case}

\par An analogous procedure starting with Eq. \eqref{A_1D} results in
\begin{equation}
A(x,z,t) = \frac{(1-i)}{2\sqrt{2\pi bz}} \int_{-\infty}^{+\infty}\text{d}x' A\left(x',0,t-\frac{e}{c}z\right)\exp[i\frac{(x-x'-a_xz)^2}{4bz}]
\label{diffraction_integral_1D}
\end{equation}

\samepage

\section{Helmholtz-Gauss beams}

\par Since $A(\zeta,\eta,z)$ satisfies Eq. \eqref{paraxial_xi_eta}, which is exactly a paraxial wave equation with $(2k_0)^{-1}$ replaced by $b$, there exists a corresponding Helmholtz-Gauss beam with envelope $A_{HG}(\zeta,\eta,z)$ given by \cite{hg_reference}
\begin{equation}
A_{HG}(\zeta,\eta,z) = \frac{1}{\mu}\exp[-q^2\frac{(\zeta^2 + \eta^2)}{\mu}]A\left(\frac{\zeta}{\mu},\frac{\eta}{\mu},\frac{z}{\mu}\right)
\end{equation}

\noindent where $\mu = 1 + i4q^2bz$ \footnote{Note that $\mu=1 + i2q^2z/k_0$ for the paraxial wave equation and we have to replace $(2k_0)^{-1}$ with $b$ to be consistent with Eq. \eqref{paraxial_xi_eta}.}.
 			
\par Using the transformation of Eq. \eqref{transformation_1}, the envelope is expressed in the original Cartesian coordinates by
\begin{equation}
A_{HG}(x,y,z) = \frac{1}{\mu}\exp\left[-\frac{q^2}{\mu\left(1-\frac{d^2}{4b^2}\right)}\left((x-a_xz)^2 +(y-a_yz)^2-\frac{d}{b}(x-a_xz)(y-a_yz)\right)\right] A\left(\frac{x}{\mu},\frac{y}{\mu},\frac{z}{\mu}\right)
\end{equation} 

\par Note that, as $\zeta$ and $\eta$ are linear combinations of $x$, $y$ and $z$, $A\left(\frac{\zeta}{\mu} , \frac{\eta}{\mu} , \frac{z}{\mu}\right)$ can be expressed as $A\left(\frac{x}{\mu} , \frac{y}{\mu} , \frac{z}{\mu}\right)$. At $z=0$, we have
\begin{equation}
A_{HG}(x,y,0) = A(x,y,0)\exp[-\frac{q^2}{\left(1-\frac{d^2}{4b^2}\right)}\left(x^2+y^2-\frac{d}{b}xy\right)]
\end{equation}

\noindent which is a Gaussian apodization with an extra mixed-product term proportional to $d$.	
	
\section{Nonparaxial linearly polarized electromagnetic field with concentrated spectra}

\par Since any Cartesian component of an electric field $\vec{E}$ obeys the scalar wave equation, a scalar wave with concentrated spectra can be assigned to one of them and the other field components can be obtained via Maxwell's equations.	

\par If we choose a linearly polarized electric field in the x-direction $\vec{E} = E_x\hat{x}+E_z\hat{z}$ with
\begin{equation}
E_x(x,y,z,t) = A(x,y,z,t) e^{-i\omega_0 t} e^{ik_{x_0}x} e^{ik_{y_0}y} e^{ik_{z_0}z} 
\end{equation}

\noindent the $E_z$ component is given by Gauss's law ($\vec{\nabla} \cdot \vec{E}=0$), which implies
\begin{equation}
E_z=-\int\frac{\partial E_x}{\partial x} \text{d}z
\end{equation}

\par Integrating by parts when $A(z,y,z,t)$ is in the integrand and using its slow variation (Eq. \eqref{inequalities_space}) to write $\int (\partial A/\partial x_i) e^{ik_{z_0}z}\,\text{d}z \approx (\partial A/\partial x_i) \int e^{ik_{z_0}z}\,\text{d}z$ (where $x_i$ can be any spatial coordinate), we find that
\begin{equation}
 E_z = e^{-i\omega_0 t} e^{ik_{x_0}x} e^{ik_{y_0}y} e^{ik_{z_0}z} \left[-a_xA +\frac{i}{k_{z_0}} \left(\frac{\partial A}{\partial x}-a_x\frac{\partial A}{\partial z}\right) \right]
\label{Ez}
\end{equation}

\par From Faraday's law ($\vec{\nabla}\times \vec{E}= - \partial \vec{B}/{\partial t}$), $\vec{B}$ can be calculated by integration in $t$. Once more, we will use integration by parts along with the slow variation of $A(x,y,z,t)$ (Eq. \eqref{inequalities_time}) to write $\int(\partial A/\partial t)e^{-i\omega_0 t} \,\text{d}t \approx (\partial A / \partial t)\int e^{-i\omega_0 t} \,\text{d}t$. Due to Eq. \eqref{inequalities}, it is also possible to neglect the second-order derivatives. Therefore, the results are
\begin{subequations}
	\begin{align}
		\nonumber B_x &= -\frac{i}{\omega_0} e^{-i\omega_0 t} e^{ik_{x_0}x} e^{ik_{y_0}y} e^{ik_{z_0}z}   \\
		& \times \left[-dk_{z_0}^2\left(iA + \frac{1}{\omega_0}\frac{\partial A}{\partial t}\right) - a_x\frac{\partial A}{\partial y} - a_y\frac{\partial A}{\partial x} + dk_{z_0}\frac{\partial A}{\partial z} \right] \\
		&\nonumber B_y = -\frac{i}{\omega_0} e^{-i\omega_0 t} e^{ik_{x_0}x} e^{ik_{y_0}y} e^{ik_{z_0}z} \\
		&\times \left[k_{z_0}\left(iA + \frac{1}{\omega_0}\frac{\partial A}{\partial t}\right)(1 + a_x^2) + \frac{\partial A}{\partial z} (1-a_x^2) + 2a_x\frac{\partial A}{\partial x} \right] \\
		&B_z = \frac{i}{\omega_0} e^{-i\omega_0 t} e^{ik_{x_0}x} e^{ik_{y_0}y} e^{ik_{z_0}z} \left[\frac{\partial A}{\partial y} + k_{y_0}\left(iA + \frac{1}{\omega_0}\frac{\partial A}{\partial t} \right) \right] 
	\end{align}
\end{subequations}

\section{Examples of Cartesian solutions}

\subsection{Gaussian pulse with Gaussian bidimensional spatial profile}

\par If we choose a separable Gaussian spectrum in $k^\prime_x$, $k^\prime_y$ and $u$:
\begin{equation}
S(k^{\prime}_x,k^{\prime}_y,u) = C \frac{r_t}{2\pi^{\frac{3}{2}}} e^{-\frac{r_x^2}{4} k_x^{\prime \, 2}} e^{-\frac{r_y^2}{4}k_y^{\prime \, 2}} e^{-\frac{r_t^2}{4}u^2}
\end{equation}

\noindent where $C$ is an arbitrary constant and $r_x$, $r_y$ and $r_t$ are related to the spot sizes in $x$, $y$ and $t$, respectively, and use the integral result $\int_{-\infty}^{+\infty}\exp(-p^2x^2 \pm qx) dx = \frac{\sqrt{\pi}}{p}\exp\left(\frac{q^2}{4p^2}\right)$, valid for $\Re(p^2) > 0$ \cite{gradstein}, Eq. \eqref{Psi_2D} and Eq. \eqref{A_2D} give
	
\begin{align}
	\nonumber&\Psi_{GB}^{2D}(x,y,z,t) = C\frac{e^{-i\omega_0 t}e^{ik_{x_0}x}e^{ik_{y_0}y}e^{i\sqrt{k^2 - k_{\mathsmaller\perp_0}^2}z}}{\sqrt{\left(\frac{r_x^2}{4} + ibz\right)\left(\frac{r_y^2}{4} + ibz\right)}\sqrt{1 + \frac{d^2z^2}{4\left(\frac{r_y^2}{4} + ibz\right)\left(\frac{r_x^2}{4} + ibz\right)}}} \exp\left[-\frac{(t - \frac{e}{c}z)^2}{r_t^2}\right]\\
	&\times \exp\left[\frac{1}{4\left(1 + \frac{d^2z^2}{4\left(\frac{r_x^2}{4} + ibz\right)\left(\frac{r_y^2}{4} + ibz\right)}\right)}\left(-\frac{\left(x - a_xz\right)^2}{\frac{r_x^2}{4} + ibz}  -\frac{\left(y - a_yz\right)^2}{\frac{r_y^2}{4} + ibz} + \frac{idz\left(x - a_xz\right)\left(y - a_yz\right)}{\left(\frac{r_x^2}{4} + ibz\right)\!\!\left(\frac{r_y^2}{4} + ibz\right)}\right)\right]
\end{align}

\subsection{Unidimensional Gaussian beam (GB)}

\par For a single frequency ($\omega_0$) and a Gaussian spectrum in $k^\prime_x$:
\begin{equation}
S(k^{\prime}_x,u) = C\delta(u)\frac{r_x}{2\sqrt{\pi}} e^{-\frac{r_x^2}{4} k_x^{\mathsmaller\prime \, 2}}
\label{gauss,1d,spectrum}
\end{equation}

\noindent the resultant wave is a unidimensional Gaussian beam (GB):
	\begin{equation}
	\Psi_{GB}^{1D} (x,z,t) \approx C \frac{e^{-i\omega_0 t} e^{i\sqrt{k_0^2-k_{\perp_0}^2}z} e^{ik_{x_0}x}}{\sqrt{1+i\frac{4b}{r_x^2}z}} \exp\left[-\frac{(x-a_x z)^2}{r_x^2\left(1+i\frac{4b}{r_x^2}z\right)}\right]
	\label{gauss,1d,beam}
	\end{equation}
		
\subsection{Nonparaxial Cartesian Parabolic-Gaussian (PG) beams}
\label{nonparaxial_cartesian_parabolic_gaussian}

\par The solutions of the paraxial wave equation known as Cartesian Parabolic-Gaussian (PG) beams were introduced by Bandres et al. in \cite{cartesian-beams} and further analyzed in \cite{parabolic-gaussian-beams}. Their envelope $_{n}A_{\nu}(x,z)$ is described by
\begin{equation}
_{n}A_{\nu}(x,z)={_{n}\Theta_{\nu}} (z) \exp\left[\frac{ik}{4}\left(\frac{1}{p}+\frac{1}{q}\right)x^2\right] {_{n}P_{\nu}}\left(\sqrt{k\left(\frac{1}{p}-\frac{1}{q}\right)}x\right)
\label{parabolic_gaussian_envelope}
\end{equation}

\noindent with
\begin{gather}
p=p_0+z \,\text{,} \quad p_0=\frac{z_R}{h+i} \\
q=q_0+z \,\text{,} \quad q_0=-p_0^* \\
{_{n}\Theta_{\nu}} (z) = 2^{1/4-n/2}\frac{(q/q_0)^{i\nu/2-1/4}}{(p/p_0)^{-i\nu/2+1/4}} \\
{_{n}P_{\nu}}(x) = \left(xe^{i\pi/4}\right)^{n-1/2}\exp\left(-\frac{ix^2}{4}\right){_{1}F_{1}}\left(\frac{n}{2}-\frac{i\nu}{2};n;\frac{ix^2}{2}\right)
\end{gather}

\noindent where $z_R=k r_0^2/2$ is the Rayleigh distance,  $r_0$ is the waist of the Gaussian envelope, $n$ is $1/2$ for even beams and $3/2$ for odd beams, $\nu\in \mathbb{R}$ is the order of the parabolic cylinder function ${_{n}P_{\nu}}(x)$, ${_{1}F_{1}}(a;b;x)$ is the confluent hypergeometric function and $h\in\mathbb{R}$ is a real parameter. At $z=0$, it becomes
\begin{equation}
_{n}A_{\nu}(x,z=0)=\frac{1}{2^{\nu/2-1/4}} \exp\left(-\frac{x^2}{r_0^2}\right) {_{n}P_{\nu}}\left(\frac{2\sqrt{h}}{r_0}x\right)
\end{equation}

\par To obtain the nonparaxial version of this envelope, is is only necessary to replace $x$ by $(x-a_x z)/\sqrt{2bk_0}$, as explained in sec. \ref{seção de redução à equação de onda}.

\par Although the previous nonparaxial examples are just inclined versions of the usual paraxial beams, they are useful for describing superpositions of beams (as commented in sec. \ref{symmetric sum of beams section}) and for generating new expressions for azimuthally symmetric waves (as shown in sec. \ref{nonparaxial_azimuthal_section} and exemplified in sec. \ref{exemplos_azimutais}). However, it is also possible to use the presented formalism to obtain waves with enhanced characteristics with respect to their paraxial counterparts, as illustrated in the next example.

\subsection{Nonparaxial Airy Beam}
		
\par The spectrum
\begin{equation}
\mathcal{S}(\tilde{k}^{\mathsmaller\prime}_x,u) = \frac{\delta(u)}{2\pi} \exp\left[i\left(\frac{\tilde{k}_x^{\mathsmaller\prime\,3}}{3} + (\beta + i\sigma)\tilde{k}_x^{\mathsmaller\prime \, 2} + \vartheta \tilde{k}_x^\mathsmaller\prime\right)\right]
\end{equation}

\noindent with free real parameters $\beta$, $\vartheta$ and $\sigma$ results in a nonparaxial Airy beam. $\sigma$ causes a Gaussian apodization of the spectrum, which is manifested in space as an exponential apodization of the Airy spatial profile at $z=0$. As in the paraxial case, the spectrum $\mathcal{S}(\tilde{k}^{\mathsmaller\prime}_x,u)$ is not concentrated unless the beam is apodized. 

\par Using the integral result $\int_{-\infty}^{+\infty} \exp\left[i\left(\frac{t^3}{3} + vt^2 + wt\right)\right]dt = 2\pi \exp\left[i(\frac{2}{3}v^3 - vw)\right] A_i(w - v^2)$ \cite{livro_Airy}, the expression of the beam becomes
\begin{align}
\hspace*{-5mm}	&\nonumber \Psi_{Ai}\,(s,\xi) = e^{-i\omega_0 t} e^{i\alpha s} e^{i\frac{\alpha^2}{\gamma^2}(1-\gamma^2)\xi} \\
\hspace*{-5mm}	& \nonumber \times A_i\left(s + \vartheta -\beta^2 + \sigma^2 -2i\beta \sigma + \xi\left(\beta + i\sigma - \alpha(1-\gamma^2)\right) - \frac{\xi^2}{4} \right) \\
\hspace*{-5mm}	& \times \exp\left\{i\left[\frac{2}{3}\left(\beta + i\sigma - \frac{\xi}{2}\right)^3  - \left(\beta + i\sigma - \frac{\xi}{2}\right)\left(s - \alpha(1-\gamma^2)\xi + \vartheta\right) \right] \right\}
\label{nonparaxial Airy beam}
\end{align}
	
\par Denoting $\delta \approx 1.01879$, the parabolic trajectory of the beam's position of maximum amplitude is
\begin{equation}
s + \vartheta -\beta^2+ \sigma^2 -2i\beta \sigma + \xi\left(\beta + i\sigma - \alpha(1-\gamma^2)\right) - \frac{\xi^2}{4}=-\delta
\end{equation}

\noindent since $x=\delta$ is where $Ai(x)$ has its maximum value. Therefore, $\beta$ controls the linear coefficient and $\vartheta$, the independent coefficient. If we choose $\sigma=0$, $\beta=\alpha(1-\gamma^2)$ and $\vartheta=-\delta+\beta^2$, the trajectory equation becomes simply
\begin{equation}
s = \frac{\xi^2}{4}
\end{equation}

\noindent or, as a function of $x$ and $z$,
	
\begin{equation}
x = \frac{1}{4k_{x_0}^2x_0^3}\frac{z^2}{(1-\gamma^2)^3}
\end{equation}

\par
\begin{figure}[h!]
\centering
\includegraphics[width=0.6\columnwidth]{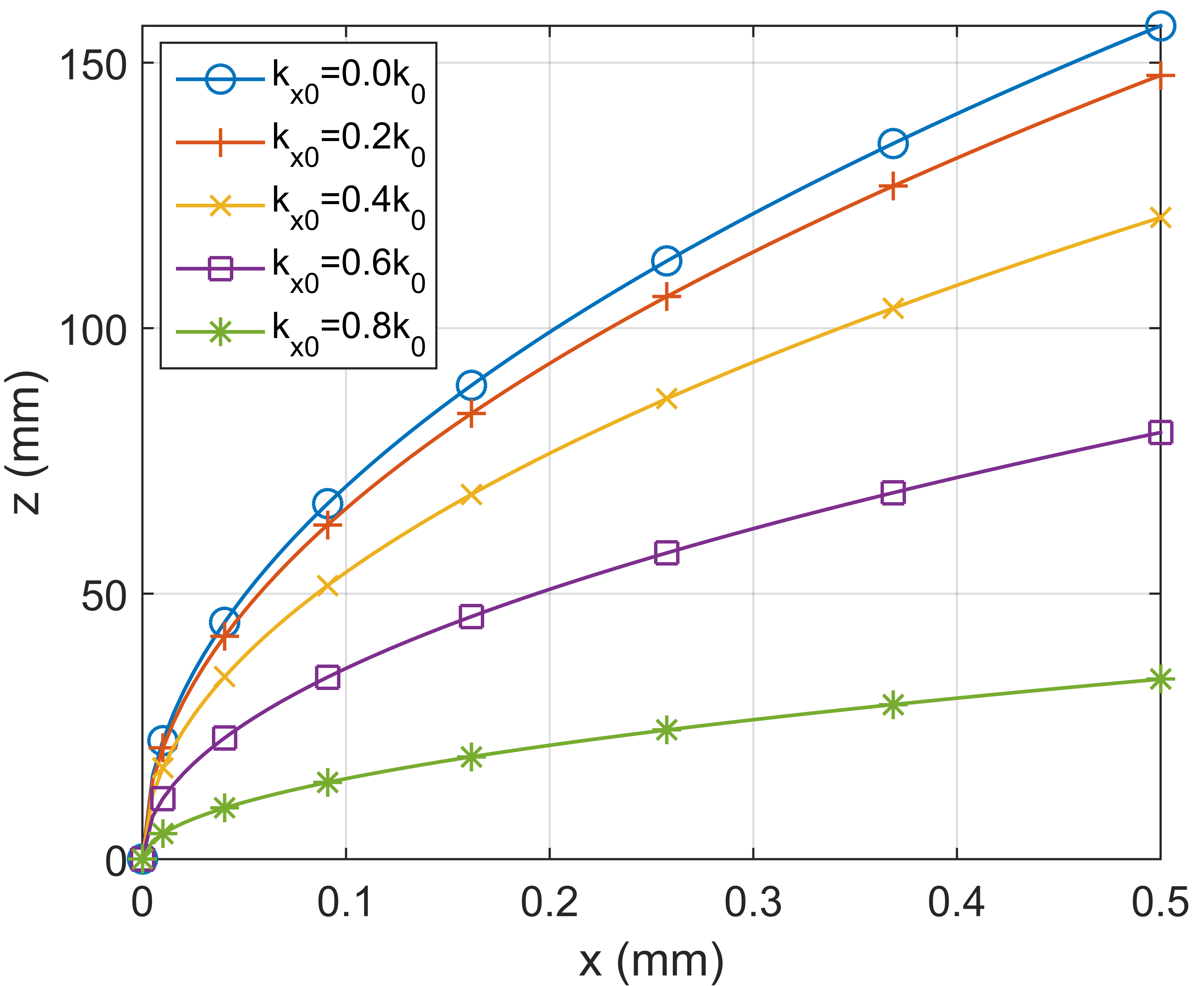}
\caption{Variation of the Airy beam's parabolic trajectory when $k_{x_0}$ varies and $x_0=50\,\mu\text{m}$.}
\label{parabolic_trajectory}
\end{figure}

 While $x_0$ controls the spot size of the beam, $k_{x_0}$ can be used to choose a desired curvature for its trajectory. To illustrate this, Fig. \ref{parabolic_trajectory} shows the variation of the beam's curvature when $k_{x_0}$ is varied for $x_0=50\,\mu\text{m}$.
				
\section{Symmetric superposition of Cartesian waves}
\label{symmetric sum of beams section}	

\par As explained in the introduction, the basic goal of the analysis so far is to provide a simple analytic way to describe the propagation of a superposition of waves traveling in different directions, thus composing interesting interference patterns in the transverse plane. Although these waves obey different approximate equations, since their dissimilar $k_{x_0}$ and $k_{y_0}$ give rise to different constants in Eq. \eqref{A_eq}, they are all approximate solutions of the homogeneous scalar wave equation, which implies that their expressions can be legitimately combined.

\par Let's take as an example the superposition of two equal unidimensional beams with the same initial envelope $A(x,z=0)$ traveling in opposite directions so that the resulting wavevector spectrum has peaks concentrated both at $+k_{x_0}$ and $-k_{x_0}$ \footnote{Note that, according to sec. \ref{seção de redução à equação de onda}, $A$ is a function of $\chi$ and $z$ and, at $z=0$, $\chi=x/\sqrt{2bk_0}$. Since $b$ is the same for $+k_{x_0}$ and $-k_{x_0}$, the initial envelope is equal for the two beams.}. Applying the diffraction integral (Eq. \eqref{diffraction_integral_1D}) for each beam and combining the results, we have
\begin{align}
	\nonumber \Psi(x,z,t) &= e^{-i\omega_0 t} e^{i\sqrt{k_0^2-k_{\mathsmaller\perp_0}^2}z} \frac{(1-i)}{2\sqrt{2\pi bz}}  \\
	&\nonumber \times\left\{e^{ik_{x_0}x} \int_{-\infty}^{+\infty} \text{d}x^{\mathsmaller\prime} A(x^{\mathsmaller\prime} , 0) \exp\left[i\frac{(x-x^{\mathsmaller\prime} - a_xz)^2}{4bz}\right]\right. \\
	& +\left. e^{-ik_{x_0}x} \int_{-\infty}^{+\infty} \text{d}x^{\mathsmaller\prime} A(x^{\mathsmaller\prime} , 0) \exp\left[i\frac{(x-x^{\mathsmaller\prime} + a_xz)^2}{4bz}\right]\right\}
\label{initial expression for symmetrical beams' sum}
\end{align}

\noindent where $a_x$ is related to $+k_{x_0}$. After expanding the squares and rearranging the terms, the result is
\begin{align}
	\nonumber \Psi(x,z,t) &= e^{-i\omega_0 t} e^{i\sqrt{k_0^2-k_{\mathsmaller\perp_0}^2}z} \frac{(1-i)}{\sqrt{2\pi bz}} \exp\left[i\frac{a_x^2z}{4b}\right]  \\
	& \times \int_{-\infty}^{+\infty} \text{d}x^{\mathsmaller\prime} A(x^{\mathsmaller\prime},0) \exp\left[i\frac{(x-x^{\mathsmaller\prime})^2}{4bz}\right] \cos\left(k_{x_0}x - \frac{a_x(x-x^{\mathsmaller\prime})}{2b}\right)
\end{align}
		
\par At the initial plane $z=0$, the simple superposition of the two patterns gives
\begin{align}
\Psi(x,z=0,t) = e^{-i\omega_0 t} \left[e^{ik_{x_0}x}A(x,0) + e^{-ik_{x_0}x} A(x,0)\right] = 2e^{-i\omega_0 t} A(x,0) \cos(k_{x_0}x)
\end{align}

\noindent showing that the slowly-varying envelope is modulated by the fast-oscillating function $\cos(k_{x_0}x)$. Therefore, the spot radius $\Delta x_0$ of the resulting beam is determined by a quarter of this cosine period, that is, $\Delta x_0=\pi/(2k_{x_0})$.

\nopagebreak

\par In any superposition, the resulting field depth $Z$ can be estimated by the distance within which the waves interfere significantly. In this simple case, we can apply a reasoning analogous to that of the upcoming sec. \ref{approximate field depth - azimuthal} to obtain $Z\approx w_0k_0/(|k_{x_0}|\sqrt{2bk_0})$, where $w_0$ is a characteristic width of the envelope $A(x,z=0)$. 
 
 \pagebreak
 
\section{Experimental results}

\par
\begin{figure}[htbp]
\centering
\includegraphics[width=0.7\columnwidth]{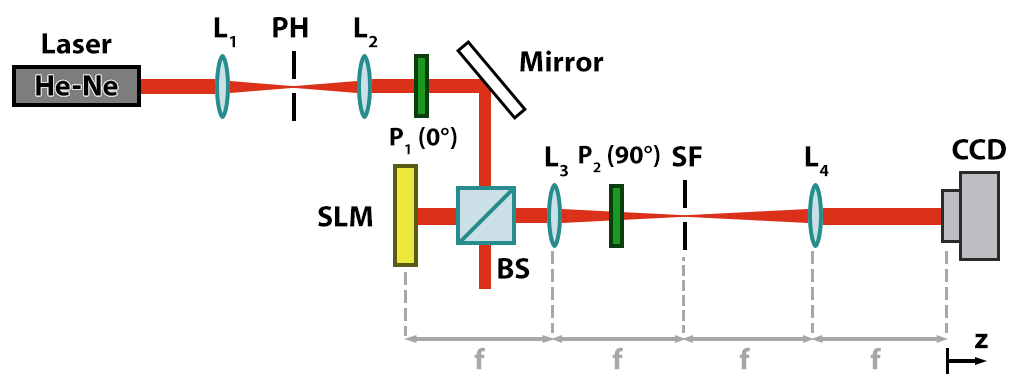}
\caption{Experimental setup for generating beams in free-space. Laser: He-Ne laser; $L1$ to $L4$: lenses; PH: pinhole; $P_1$: polarizer; BS: beam splitter; SLM: reflective Spatial Light Modulator, model LC-R1080 from Holoeye Photonics; $P_2$: analyzer; SF: circular pupil; CCD: CCD camera model DMK 41BU02.H from The Imaging Source.}
\label{experimental_setup}
\end{figure}

 To demonstrate that the proposed theoretical expressions can be used to easily describe a superposition of beams with concentrated spectra propagating in different directions, we experimentally generated two superpositions of Gaussian beams that result in illustrative interference patterns. The experimental setup is shown in Fig. \ref{experimental_setup} and used a reflective Spatial Light Modulator (SLM) to generate the desired patterns.

\par A He-Ne laser ($\lambda \approx 632.8\,\text{nm}$) is expanded and collimated by a system of lenses ($L_1$ and $L_2$) and then sent to the SLM, which is a LC-R1080 model from Holoeye Photonics and possesses a display matrix of 1920x1200 with $8.1\,\mu\text{m}$ pixel size. The device was used in amplitude modulation mode, with the polarizer and analyzer angles ($0^\circ$ and $90^\circ$, respectively) measured with respect to the SLM axis. The reflected beam is then subject to a 4f spatial filtering system. The SLM is positioned at the input plane (at a focal distance $f$ behind the lens $L_3$) and the spatial filtering mask (SF, a band-pass circular pupil) is placed at the Fourier plane to select the first diffraction order containing the information of the desired complex field pattern. After another Fourier transform, the field is obtained at the Fourier plane of lens $L_4$ and its intensity profile for different propagation distances $z$ is measured by a CCD camera (DMK 41BU02.H from The Imaging Source, with a 1280x960 display and $4.65\,\mu\text{m}$ pixel size).

\par Due to the size of the SLM's pixel, it cannot resolve well the fast amplitude and phase variations of beams with $k_{x_0}/k_0$ and/or $k_{y_0}/k_0$ ratios that are not small, so we had to choose close-to-zero values for them. Even though these beams are not significantly nonparaxial, the results illustrate how interesting interference patterns can be produced. With suitable equipment, it is straightforward to generate the same kinds of beams in a nonparaxial regime.

\subsection{Superposition of two Gaussian beams}
\label{exp_double_gauss}

\par
\begin{figure}[h!]
\centering
\includegraphics[width=0.7\columnwidth]{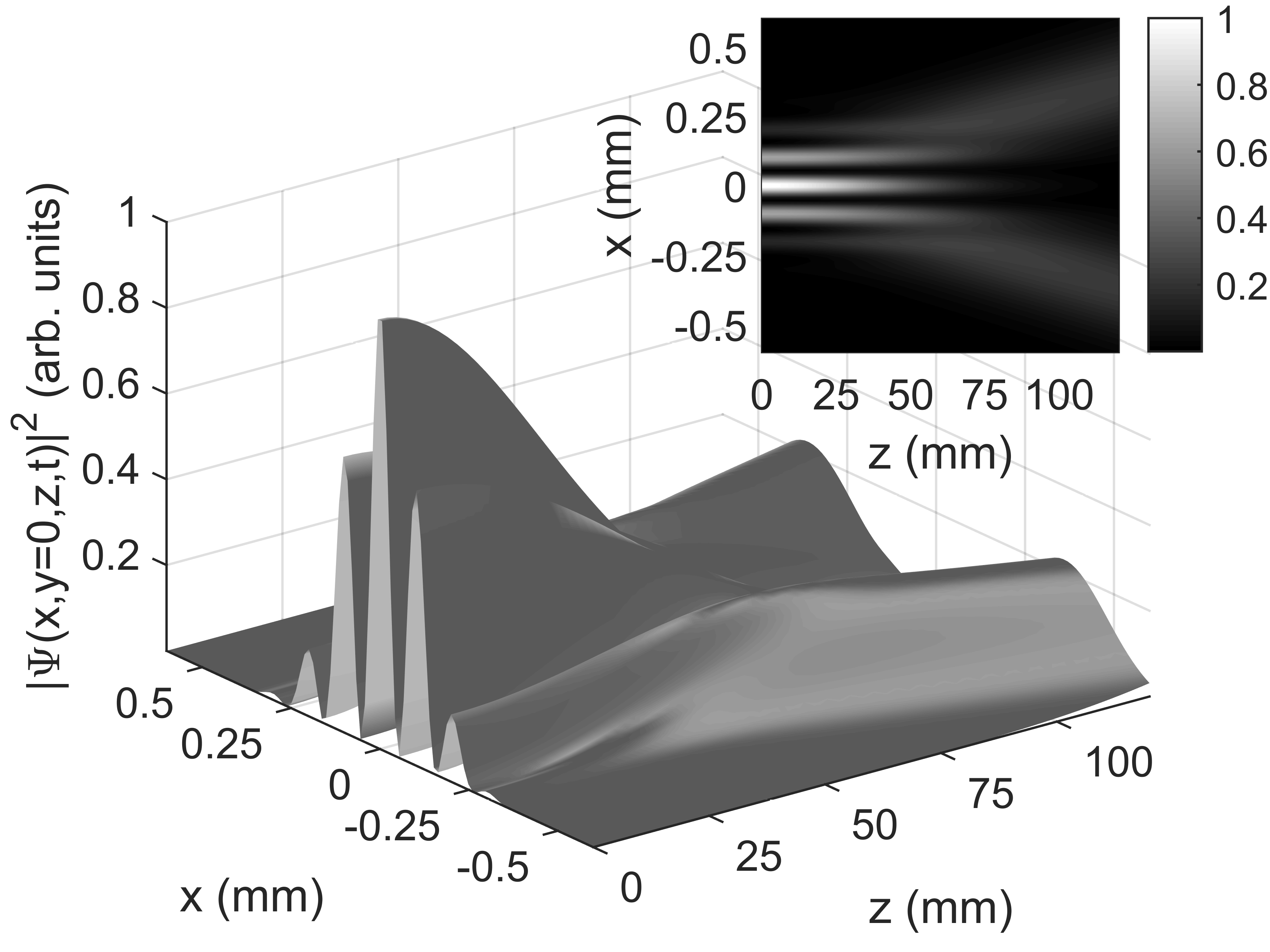}
\caption{Superposition of two Gaussian beams: theoretical intensity pattern for $y=0$.}
\label{double_gauss_cut_teorico}
\end{figure}

 The first example consists in a superposition of two Gaussian beams with same amplitudes and spot sizes $r_x=r_y=300/\sqrt{2}\,\mu\text{m}$, but with different values of $k_{x_0}$ and $k_{y_0}$. One of them has $(k_{x_0},k_{y_0})=(+0.003k_0,0k_0)$ and the other has $(k_{x_0},k_{y_0})=(-0.003k_0,0k_0)$, so that they propagate in opposite directions in the $x$-axis but do not shift in the $y$-direction. Fig. \ref{double_gauss_cut_teorico} shows what we should expect for a $y=0$ cut of the beam's intensity profile: a sinusoidal pattern enveloped by a Gaussian function. Therefore, it can be viewed as a kind of ``cartesian analog'' of a Bessel-Gauss (BG) beam, which consists in a superposition of GBs over a cone and has a Bessel function profile enveloped by a Gaussian function.

\par
\begin{figure}[h!]
\centering
\includegraphics[width=0.7\columnwidth]{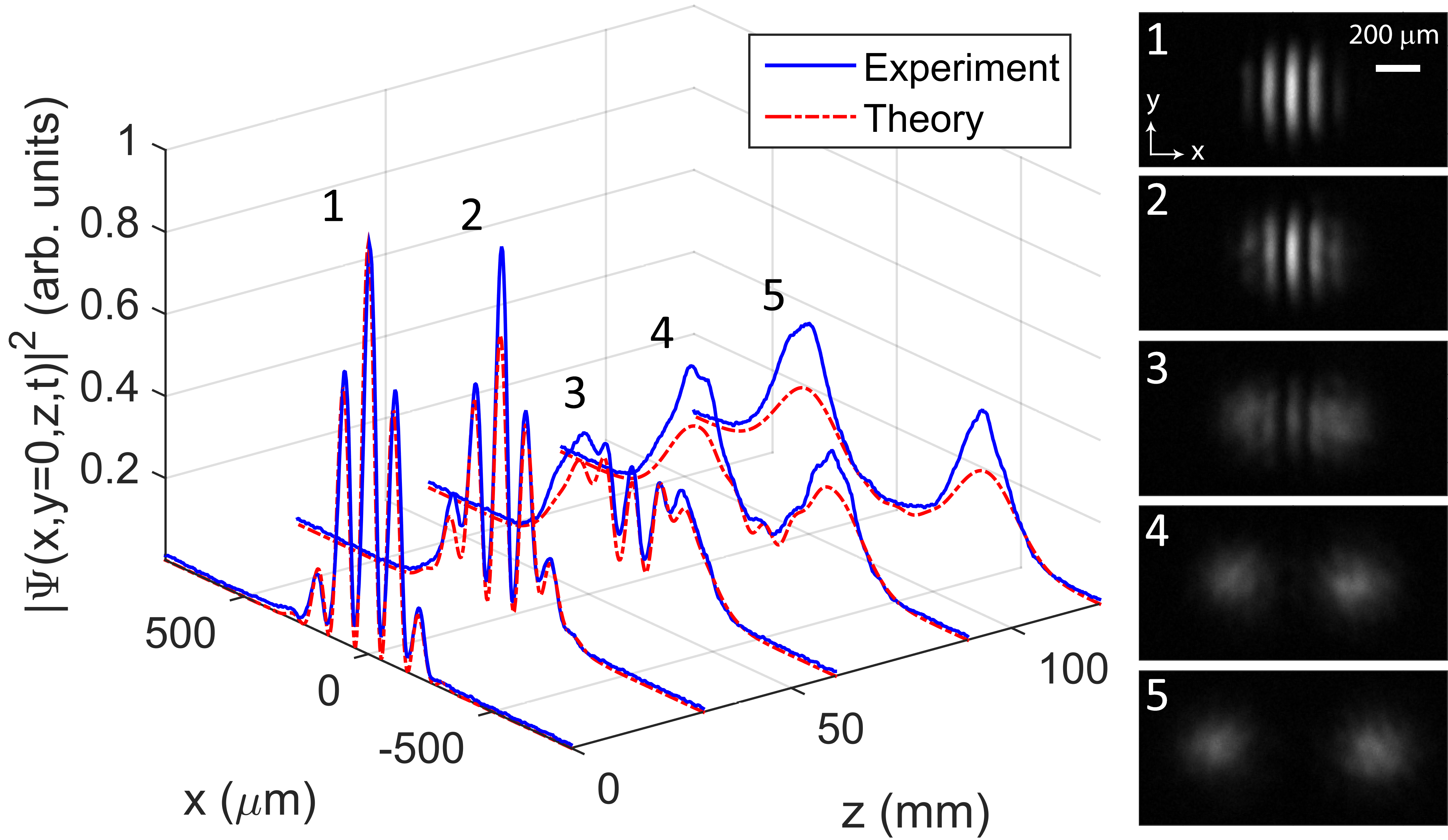}
\caption{Superposition of two Gaussian beams: comparison between the experimental measurements and the theoretical predictions for $y=0$ cuts at different propagation distances and captured patterns for each of them.}
\label{double_gauss_cut_exp}
\end{figure}

 The propagation distance $Z$ for which the sinusoidal pattern is kept can be estimated by the expression in sec. \ref{symmetric sum of beams section}, resulting in $Z\approx r_x k_0/(|k_{x_0}|\sqrt{2bk_0}) \approx 70.71\,\text{mm}$, which agrees with Fig. \ref{double_gauss_cut_teorico}. Fig. \ref{double_gauss_cut_exp} shows the good agreement between the theoretical and the measured intensity patterns for $y=0$ cuts at different propagation distances and also presents the measured patterns for each of these positions.

\subsection{Superposition of four Gaussian beams} 
\label{exp_double_double_gauss}

\par The second example consists in a superposition of four Gaussian beams with same amplitudes and spot sizes $r_x=r_y=300/\sqrt{2}\,\mu\text{m}$, but with different values of $k_{x_0}$ and $k_{y_0}$. Two of them are the same as the ones in sec. \ref{exp_double_gauss} and the other two have $(k_{x_0},k_{y_0})=(0k,+0.003k)$ and $(k_{x_0},k_{y_0})=(0k,-0.003k)$, so that, in addition to the two beams being shifted in the x-direction, we now have two more beams that are shifted along the y-axis in opposite directions. Therefore, the interference pattern will be a little more sophisticated, with more maxima and minima than the one in sec. \ref{exp_double_gauss}.

\begin{figure}[htbp]
\centering
\includegraphics[width=0.7\columnwidth]{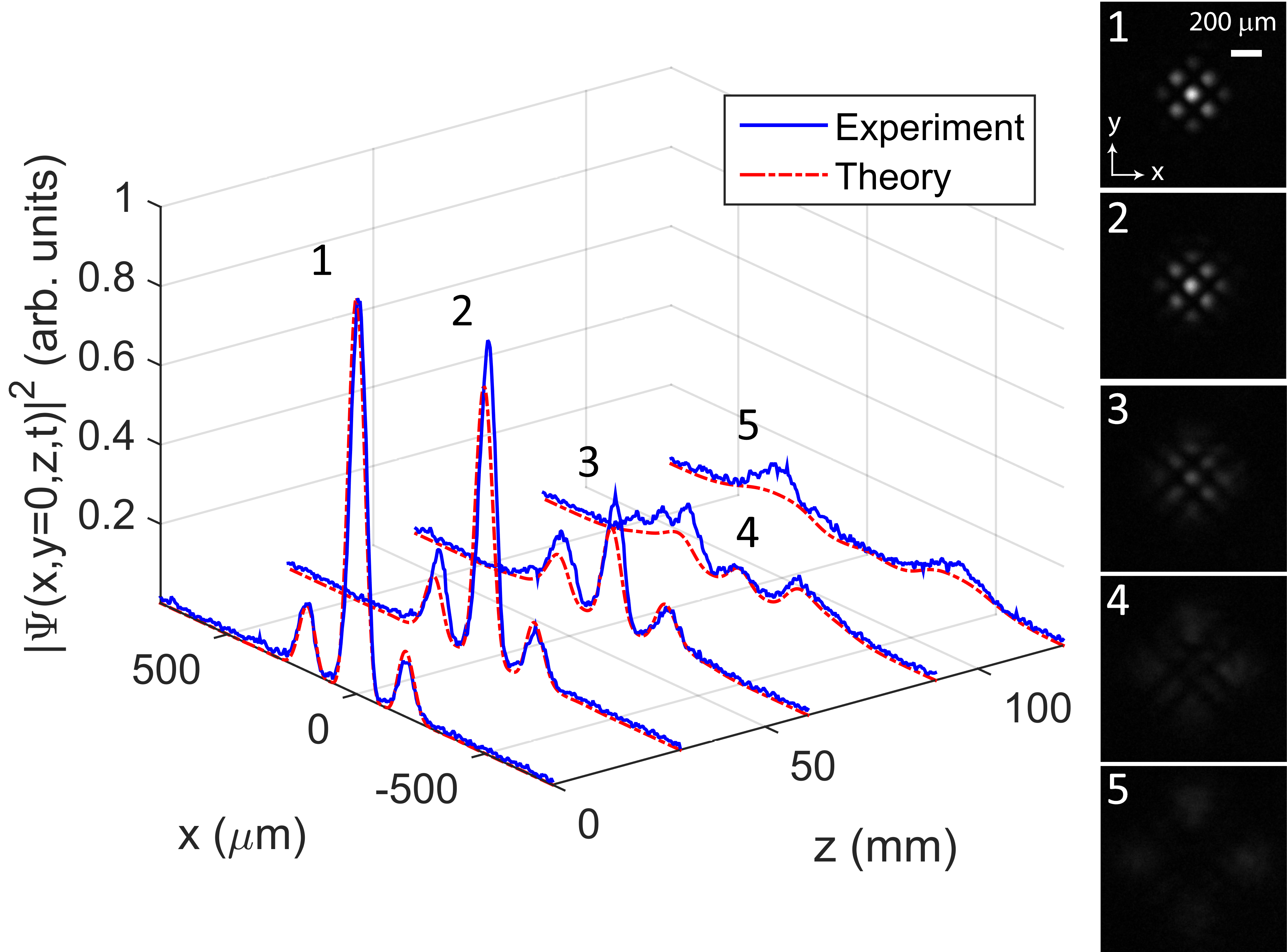}
\caption{Superposition of four Gaussian beams: comparison between the experimental measurements and the theoretical predictions for $y=0$ cuts at different propagation distances and captured patterns for each of them.}
\label{double_double_gauss_cut_exp}
\end{figure}

Fig. \ref{double_double_gauss_cut_exp} shows the good agreement between the theoretical and measured intensity patterns for $y=0$ cuts at different propagation distances and also presents the measured patterns for each of these positions.

\section{Nonparaxial azimuthally symmetric waves with concentrated spectra}
\label{nonparaxial_azimuthal_section}

\par Following a procedure similar to the one in sec. \ref{integral_formulation_and_PDE}, we now present how the previous unidimensional results can be leveraged to describe nonparaxial azimuthally symmetric waves with concentrated spectra.

\subsection{Superposition of Bessel beams with close cone angles}
	
\par Any azimuthally symmetric wave $\Psi(\rho,z,t)$ can be expressed as a superposition of Bessel beams of order zero, that is
\begin{equation}
\Psi(\rho,z,t) = \int_{-\infty}^{+\infty}\text{d}\omega\, e^{-i\omega t}\int_{0}^{+\infty}\text{d}k_\rho\,\tilde{S}(k_\rho,\omega) J_0(k_\rho \rho)e^{i\sqrt{k^2-k_\rho^2}z}
\label{azimuthal_exato}
\end{equation}
	
\noindent with the spectrum $\tilde{S}(k_\rho,\omega)$ defining the amplitudes of each beam. Note that the upper limit of $k_\rho$ is not limited to $\omega/c$, so that evanescent waves are allowed.

\nopagebreak

\par For an arbitrary spectrum, the integrals in Eq. \eqref{azimuthal_exato} generally do not have a closed analytical representation, but when $\tilde{S}(k_\rho,\omega)$ is concentrated around $(k_\rho,\omega)=(k_{\rho_0},\omega_0)$ some simplifications can be applied using an approximate expression for $J_0(\cdot)$. As in sec. \ref{bidimensional}, a concentrated spectrum means that $|\tilde{S}(k_\rho,\omega)|$ is significant only inside the region $|k_\rho^\prime/k_{\rho_0}|\ll 1$ and $|u/\omega_0|\ll 1$, where $k_\rho^\prime\myeq k_\rho-k_{\rho_0}$ and $u \myeq \omega-\omega_0$. In other words, all the superposed  Bessel beams have almost the same cone angle. Note, however, that it can be as large as desired, unlike in the paraxial approximation, which restricts the angle to small values.

\par Similarly to what was done in sec. \ref{integral_formulation_and_PDE}, the hypothesis of concentrated spectrum allows us to accurately approximate $k_z=\sqrt{k^2-k_\rho^2}$ to second order in $k_{\rho}^{\prime}$ and to first order in $u$:
\begin{equation}
\sqrt{k^2-k_\rho^2} \approx \sqrt{k_0^2-k_{\rho_0}^2} - ak_{\rho}^{\prime} -bk_{\rho}^{\prime \, 2} + \frac{e}{c}u
\end{equation}

\pagebreak

\par In this expression, the constants are defined as:

\begin{subequations}
\begin{align}
a \myeq \frac{k_{\rho_0}}{\sqrt{k_0^2-k_{\rho_0}^2}}\\
b \myeq \frac{1}{2}\frac{k_0^2}{(k_0^2-k_{\rho_0}^2)^{\frac{3}{2}}}\\
e \myeq \frac{k_0}{\sqrt{k_0^2-k_{\rho_0}^2}}
\end{align}
\end{subequations}

\par Changing the integration variable in Eq. \eqref{azimuthal_exato} from $k_\rho$ to $k_\rho^{\prime}$, the result is
\begin{equation}
\Psi(\rho,z,t) \approx e^{-i\omega_0 t}e^{\sqrt{k_0^2 - k_{\rho_0}^2}z}\int_{\infty}^{+\infty} \text{d}u \, e^{-iut} \int_{-k_{\rho_0}}^{+\infty} \text{d}k_\rho^{\prime}\, S(k_\rho^{\prime},u) J_0[(k_\rho^{\prime} + k_{\rho_0})\rho]e^{ik_z^{\prime}z}
\label{azimuthally symmetric approximate expression}
\end{equation}

\noindent with $S(k_\rho^{\prime},u) \myeq \tilde{S}(k_\rho^{\prime} + k_{\rho_0}, u + \omega_0)$ and 
\begin{equation}
k_z^{\prime} = -ak_\rho^{\prime} - bk_{\rho}^{\prime \, 2} + \frac{e}{c}u
\end{equation}

\par For $k_{\rho_0}\neq 0$, the spectrum is concentrated around $k_\rho^{\prime} = 0$ and its amplitude is negligible in the region $\abs{k_\rho^{\prime}} \geq k_{\rho_0}$. Therefore, the lower limit of the integral in $k_\rho^\prime$ in Eq. \eqref{azimuthally symmetric approximate expression} can be extended to $-\infty$ without loss of accuracy.

\subsection{An approximate expression of \boldmath$J_0(x)$}

\par Now, we will make use of an unusual, although very accurate, approximate expression for $J_0(x)$. It was first introduced in \cite{roger}, but the reasoning behind its derivation was not presented. Therefore, we show here how this approximation is obtained.

\par The asymptotic approximation for $J_0(x)$ is \cite{bessel}
\begin{equation}
J_0(x) \approx \sqrt{\frac{2}{\pi x}}\cos\left(x-\frac{\pi}{4}\right)
\label{bessel_asymp}
\end{equation}

\par
\begin{figure}[h!]
\centering
\includegraphics[width=0.6\columnwidth]{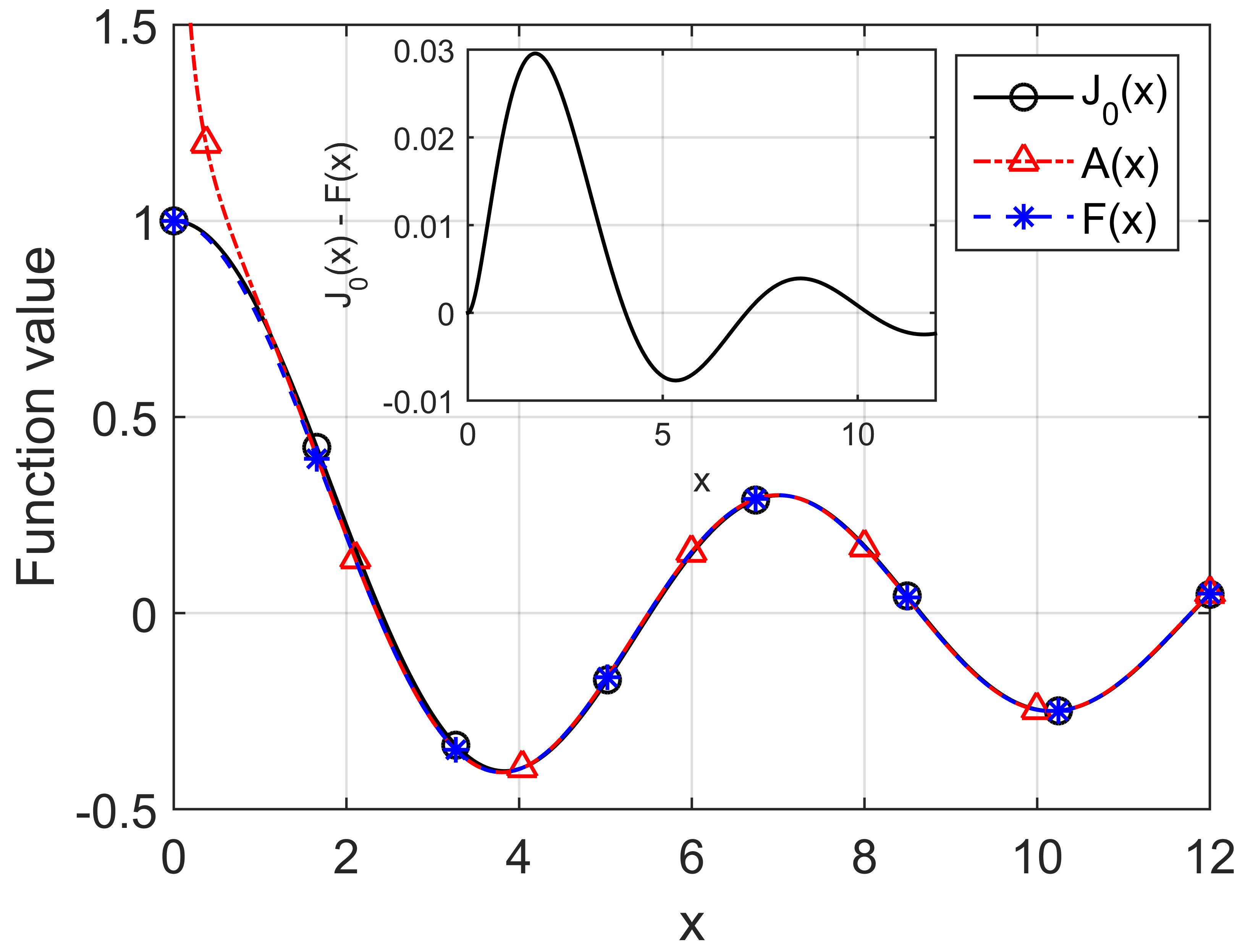}
\caption{Comparison among $J_0(x)$, its asymptotic approximation $A(x)$ (Eq. \eqref{bessel_asymp}) and the unusual approximation $F(x)$ (Eq. \eqref{bessel_approx_expression}). The inset shows the difference between $J_0(x)$ and $F(x)$.}
\label{bessel_approx}
\end{figure}

 Although it is rigorously valid only when the argument is large, Fig. \ref{bessel_approx} shows that this asymptotic approximation is precise even for relatively small values of $x$, failing only due to the divergence when $x \rightarrow 0$. Therefore, one could attempt to improve the accuracy of Eq. \eqref{bessel_asymp} for small arguments by introducing a function in the denominator of the square root to eliminate the divergence, that is,
\begin{equation}
F(x) \approx \sqrt{\frac{2}{\pi x + g(x)}}\cos\left(x-\frac{\pi}{4}\right)
\end{equation}

\noindent where $g(x)$ is chosen in such a way that $F(x) \approx J_0(x)$ for all $x$. To this end, $g(x)$ must have the following properties:
\begin{itemize}
\item $g(0)=1$, so that $F(0)=J_0(0)=1$.
\item $g^\prime(0)=-(\pi-2)$, so that $F^\prime(0)=J_0^\prime(0)=0$.
\item $g(x) \rightarrow 0$ sufficiently fast when $x$ increases, so that $F(x) \approx J_0(x)$ for $x \neq 0$.
\end{itemize}

\par The simplest function that satisfies all these conditions is $g(x)=e^{-(\pi-2)x}$ and we can hence accurately approximate $J_0(x)$ with the expression
\begin{equation}
J_0(x) \approx \sqrt{\frac{2}{\pi x + e^{-(\pi-2)x}}}\cos\left(x-\frac{\pi}{4}\right)
\label{bessel_approx_expression}
\end{equation} 

\noindent for any $x \geq 0$, even small. To allow negative arguments, $x$ in the right-hand side of Eq. \eqref{bessel_approx_expression} should be replaced by $|x|$, so that the expression is even, as is $J_0(x)$. Fig. \ref{bessel_approx} displays a comparison between the exact $J_0(x)$, the asymptotic approximation and the unusual approximation, showing that the latter represents $J_0(x)$ very accurately.

\par It is worth noting that the approximation in Eq. \eqref{bessel_approx_expression} for $J_0(x)$ separates the slowly-decaying envelope $\sqrt{\frac{2}{\pi x + e^{-(\pi-2)x}}}$ from the fast oscillations expressed by $\cos\left(x-\frac{\pi}{4}\right)$.

\subsection{Similarities with the unidimensional Cartesian case}
\label{similarities with cartesian case section}

\par Extending the lower limit of the integral in Eq. \eqref{azimuthally symmetric approximate expression} to $-\infty$ and using the approximation for the Bessel function, the expression becomes
\begin{align}
\nonumber \Psi(\rho,z,t) \approx & e^{-i\omega_0 t}e^{\sqrt{k_0^2 - k_{\rho_0}^2}z}\int_{-\infty}^{+\infty} \text{d}u \, e^{-iut}\int_{-\infty}^{+\infty} \text{d}k_\rho^{\prime}\, S(k_\rho^{\prime},u) \\
&\times \sqrt{\frac{2}{\pi (k_\rho^\prime+k_{\rho_0}) \rho + e^{-(\pi-2)(k_\rho^\prime+k_{\rho_0})\rho}}} \cos\left[(k_\rho^\prime+k_{\rho_0})\rho-\frac{\pi}{4}\right] e^{ik_z^{\prime}z}
\label{azimuthally symmetric approximate expression with approximate bessel function}
\end{align}

\par Since by hypothesis $|k_\rho^\prime| \ll k_{\rho_0}$ where $|S(k_\rho^{\prime},u)|$ is significant, the argument $(k_\rho^\prime+k_{\rho_0})\rho \approx k_{\rho_0}\rho$ of the Bessel function is always positive in this region, so there is no need to include the absolute value for the approximate expression to be valid. The effect of the small variation $k_\rho^\prime$ is significant in the fast oscillations of the cosine, but can be neglected in the envelope, which remains always approximately equal to $\sqrt{\frac{2}{\pi k_{\rho_0} \rho + e^{-(\pi-2)k_{\rho_0}\rho}}}$ and can be taken outside the integral. Expanding the cosine into complex exponentials, we finally get
\begin{align}
\Psi(\rho,z,t) \approx \frac {e^{-i\omega_0t}e^{i\sqrt{k_0^2-k_{\rho_0}^2}z}}{\sqrt{2(\pi k_{\rho_0}\rho+ e^{-(\pi-2)k_{\rho_0}\rho})}} \left[e^{i(k_{\rho_0}\rho-\frac{\pi}{4})}A(\rho,z,t)+e^{-i(k_{\rho_0}\rho-\frac{\pi}{4})}A(-\rho,z,t)\right]
\label{Azimuthally symmetric pattern as a function of A}
\end{align}

\noindent where

\begin{align}
A(\rho,z,t) \myeq \int_{-\infty}^{+\infty} \text{d}u \int_{-\infty}^{+\infty}\text{d}k^{\prime}_\rho\, S(k^{\prime}_\rho,u)e^{-iut}e^{ik^{\prime}_\rho\rho}e^{ik^{\prime}_z z}
\label{Definition of A for azimuthally symmetric case}
\end{align}

\noindent has the same functional form of a unidimensional nonparaxial Cartesian envelope (Eq. \eqref{A_1D}) with $x$ replaced by $\rho$. The differential equation satisfied by $A(\rho,z,t)$ is then
\begin{align}
\frac{\partial A}{\partial z} = -\frac{e}{c}\frac{\partial A}{\partial t} - a\frac{\partial A}{\partial \rho} + ib\frac{\partial^2 A}{\partial \rho ^2}
\end{align}

\noindent and the unidimensional transformations of sec. \ref{seção de redução à equação de onda} are applicable if $x$ is replaced by $\rho$.

\par Therefore, the azimuthally symmetric version of a nonparaxial Cartesian wave can be obtained by simply applying the known solution $A(x,z,t)$ to Eq. \eqref{Azimuthally symmetric pattern as a function of A} with $x$ replaced by $\rho$ and $k_{x_0}$, by $k_{\rho_0}$. The minus sign in $A(-\rho,z,t)$ can be viewed alternativelly as a change in the sign of $a$, implying that in Eq. \eqref{Azimuthally symmetric pattern as a function of A} there are waves traveling in opposite directions.

\par Over the axis ($\rho=0$), the wave is 
\begin{equation}
\Psi(0,z,t)\approx e^{-i\omega_0t}e^{i\sqrt{k_0^2-k_{\rho_0}^2}z} A(0,z,t)
\label{on-axis-intensity-azymuthal}
\end{equation}

\noindent that is, $|\Psi(0,z,t)|=|A(0,z,t)|$, thus allowing the easy prediction of the on-axis intensity. On the other hand, at $z=0$, the result is
\begin{equation}
\Psi(\rho,0,t) \approx \frac {e^{-i\omega_0t}\left[e^{i(k_{\rho_0}\rho-\frac{\pi}{4})}A(\rho,0,t)+e^{-i(k_{\rho_0}\rho-\frac{\pi}{4})}A(-\rho,0,t)\right]}{\sqrt{2(\pi k_{\rho_0}\rho+ e^{-(\pi-2)k_{\rho_0}\rho})}} 
\label{Azimuthally symmetric wave at z=0}
\end{equation}

\par If $A(\rho,0,t)$ has a definite parity in $\rho$, the numerator of Eq. \eqref{Azimuthally symmetric wave at z=0} becomes $2\cos(k_{\rho_0}\rho-\pi/4)A(\rho,0,t)$ if $A$ is even and $2i\sin(k_{\rho_0}\rho-\pi/4)A(\rho,0,t)$ if $A$ is odd, showing that the slowly-varying envelope is modulated by a fast-oscillating sine or cosine factor with period $2\pi/k_{\rho_0}$, as in sec. \ref{symmetric sum of beams section}. In these cases, we may expect to have waves with spot radii of the order of $\Delta \rho_0\sim \pi/(2k_{\rho_0})$.

\subsection{Approximate field depth}
\label{approximate field depth - azimuthal}

\par Since the wave in Eq. \eqref{Azimuthally symmetric pattern as a function of A} can be qualitatively viewed as a superposition of waves with envelope $A(\rho,z,t)$ over a cone of half-angle $\theta=\arcsin\left(k_{\rho_0}/k\right)$, we can estimate its field depth $Z$ as the distance within which their interference is significant. From \cite{gori}, we know that this interpretation, besides being qualitatively appealing, is also quantitatively accurate for determining the field depth of paraxial waves, so we can use the unidimensional transformations of sec. \ref{seção de redução à equação de onda} to generalize this result to the nonparaxial case considered here. 

\par Let $w_0$ be a characteristic width of $A(\rho,z,t)$ at $z=0$ which embraces its main features, such as its spot radius, and let's assume its change is negligible when propagating the distance $Z$. This is reasonable because $A(\rho,z,t)$ is a slowly-varying envelope by hypothesis. In the paraxial case, $Z$ can be estimated by $Z\approx w_0/\sin\theta=w_0k_0/k_{\rho_0}$ \cite{gori}. Since the expression for the nonparaxial envelope $A(\rho,z,t)$ is obtained from an initial paraxial expression by replacing $\rho$ by $(\rho-a_x z)/\sqrt{2bk_0}$ (see sec. \ref{seção de redução à equação de onda}), if the characteristic width $w_0$ of $A(\rho,z,t)$ is converted back to a paraxial counterpart of this envelope, the result is a characteristic width of $w_0/\sqrt{2bk_0}$. Therefore, combining this converted length with the paraxial approximation for $Z$, we get
\begin{equation}
Z\approx \frac{w_0/\sqrt{2bk_0}}{\sin\theta}=\frac{w_0}{\sqrt{2bk_0}}\frac{k_0}{k_{\rho_0}} 
\label{field_depth_expression}
\end{equation}

\noindent which is an accurate estimate for the field depth of any azimuthally symmetric wave with concentrated spectra, as illustrated in the examples of sec. \ref{exemplos_azimutais}.

\section{Examples of nonparaxial azimuthally symmetric beams}	
\label{exemplos_azimutais}
	
\subsection{Bessel-Gauss (BG) beam}

\par A nonparaxial Bessel-Gauss (BG) beam is obtained by choosing a Gaussian spectrum concentrated around $k_{\rho_0}$, as the one in Eq. \eqref{gauss,1d,spectrum} with $k_x^\prime$ replaced by $k_\rho^\prime$ and $r_x$, by $r_0$. Using the Cartesian expression of Eq. \eqref{gauss,1d,beam} in Eq. \eqref{Azimuthally symmetric pattern as a function of A}, the result is	
	\begin{align}
	&\nonumber\Psi_{BG}(\rho,z,t) = \frac{e^{-i\omega_0 t} e^{i\sqrt{k_0^2-k_{\rho_0}^2}z}}{\sqrt{2\left(\pi k_{\rho_0}\rho + e^{-(\pi-2)k_{\rho_0}\rho}\right)}} \frac{C}{\sqrt{1+i\frac{4b}{r_0^2}z}} \\
	& \times \left[e^{i\left(k_{\rho_0}\rho-\frac{\pi}{4}\right)} \exp\left[-\frac{(\rho-az)^2}{r_0^2\left(1+i\frac{4bz}{r_0^2}\right)}\right] + e^{-i\left(k_{\rho_0}\rho-\frac{\pi}{4}\right)} \exp\left[-\frac{(\rho+az)^2}{r_0^2\left(1+i\frac{4bz}{r_0^2}\right)}\right]\right]
	\label{bessel gauss - our expression}
	\end{align}
	
\par At $z=0$, the beam's pattern is a Bessel function $J_0(k_{\rho_0}\rho)$ (in the form of Eq. \eqref{bessel_approx_expression}) with spot size $\Delta \rho_0 \approx 2.4/k_{\rho_0}$ apodized by a Gaussian function of waist $r_0$. 

\par For practical situations, it is desired to have $r_0 \gg \Delta \rho_0$, so that a large number of the Bessel's rings are unattenuated and the BG beam behaves like a Bessel beam for longer distances, resulting in a higher field depth, which can be estimated by making $w_0=r_0$ in Eq. \eqref{field_depth_expression}. Note that a large $r_0$ implies a concentrated spectrum, which is consistent with our assumptions.

\par For comparison, the usual expression for a BG beam, valid for paraxial regime, is \cite{gori}
\begin{align}
\Psi_{BG}^{paraxial}(\rho,z,t)=-e^{-i\omega_0 t}\frac{ik_0 C}{2zQ}\exp\left[ik_0 \left(z+\frac{\rho^2}{2z}\right)\right] J_0\left(\frac{ik_{\rho_0}k_0\rho}{2zQ}\right) \exp\left[-\frac{1}{4Q}\left(k_{\rho_0}^2+k_0^2\frac{\rho^2}{z^2}\right)\right]
\label{BG_gori}
\end{align} 

\noindent with $Q=1/r_0^2-ik_0/(2z)$.

\par
\begin{figure}[htbp]
\centering
\includegraphics[width=0.7\columnwidth]{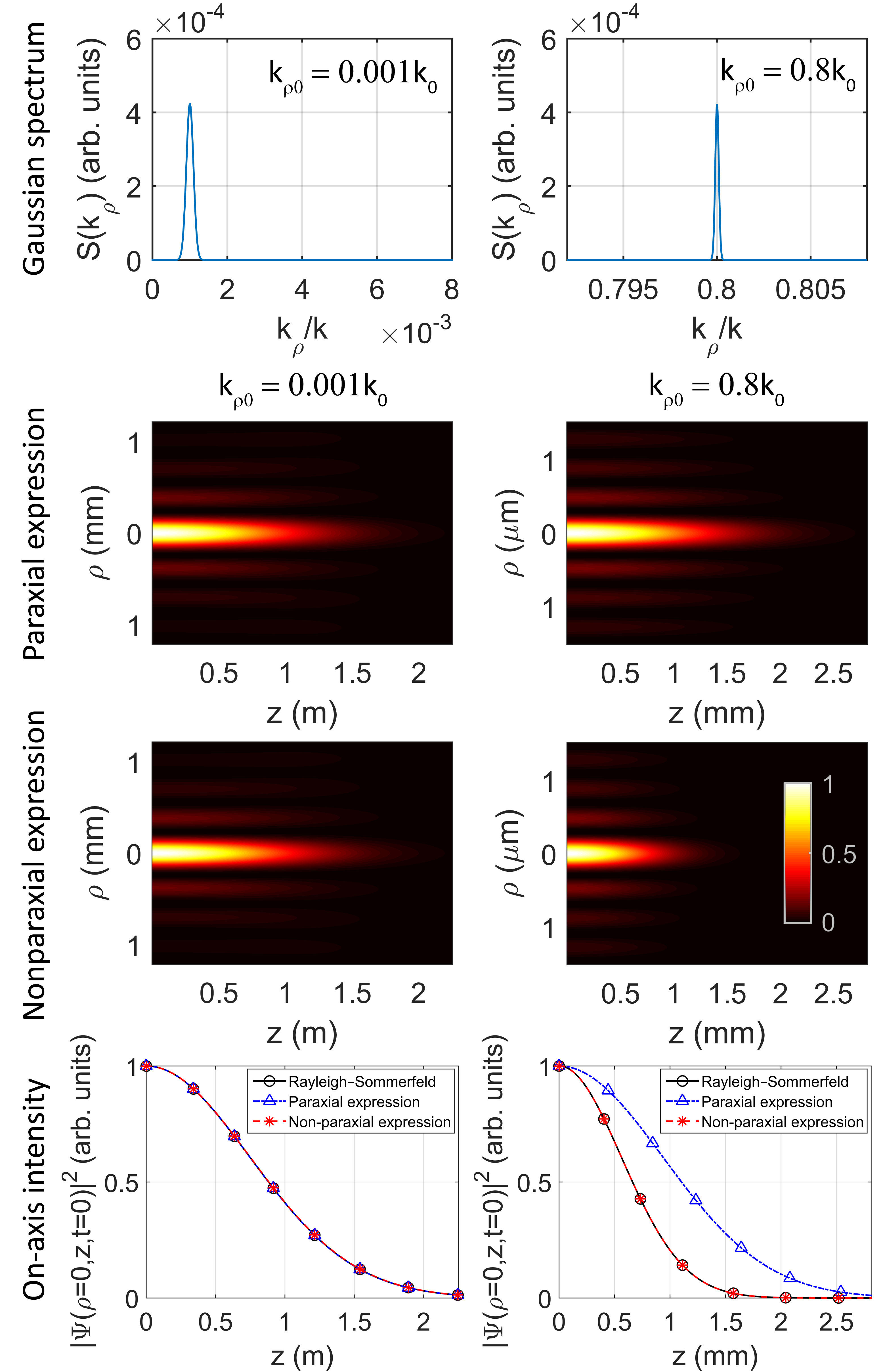}
\caption{Comparison between the paraxial (Eq. \eqref{BG_gori}) and generalized nonparaxial (Eq. \eqref{bessel gauss - our expression}) expressions for a BG beam with $r_0=1500\,\mu\text{m}$ propagating in air under paraxial ($k_{\rho_0}=0.001k_0$) and nonparaxial ($k_{\rho_0}=0.8k_0$) regimes. In both cases, $\lambda=632.8\,\text{nm}$. The first line shows the spectra, the following two lines depict the $|\Psi(\rho,z,t)|^2$ profiles according to the expressions and the last line compares the predicted on-axis intensities with the correct results calculated via Rayleigh-Sommerfeld diffraction integral.}
\label{bessel-gauss}
\end{figure}

 Fig. \ref{bessel-gauss} contrasts Eq. \eqref{BG_gori} and Eq. \eqref{bessel gauss - our expression} for two situations: paraxial ($k_{\rho_0}=0.001k_0$) and highly nonparaxial ($k_{\rho_0}=0.8k_0$). In both cases, the medium is air, $\lambda=632.8\,\text{nm}$ and $r_0=1500\,\mu\text{m}$. The first line of Fig. \ref{bessel-gauss} depicts the Gaussian spectra, showing that they are very concentrated. The following two lines show the intensities $|\Psi(\rho,z,t)|^2$ predicted by the paraxial expression (Eq. \eqref{BG_gori}, second line) and the generalized nonparaxial expression (Eq. \eqref{bessel gauss - our expression}, third line) \footnote{Note that the intensities were normalized by their peak at $z=0$.}. It is clear that the two expressions agree in the paraxial regime, but in the nonparaxial case Eq. \eqref{BG_gori} overestimates the field depth predicted by Eq. \eqref{bessel gauss - our expression}. The last line shows the intensities at $\rho=0$ predicted by both expressions and compares them with the exact intensity calculated using the Rayleigh-Sommerfeld diffraction integral for the initial pattern $\Psi(\rho,0,t)=e^{-i\omega_0 t}J_0(k_{\rho_0}\rho)\exp(-\rho^2/r_0^2)$. The three results coincide for the paraxial regime, but only Eq. \eqref{bessel gauss - our expression} agrees with the diffraction integral for the nonparaxial situation. According to Eq. \eqref{field_depth_expression}, the estimated field depths are $Z\approx 1.5\,\text{m}$ and $Z\approx 0.87\,\text{mm}$ for the paraxial and nonparaxial examples, respectively, which accurately match the diffraction integral results.

\par It is worth mentioning that Eq. \eqref{bessel gauss - our expression}, while valid for the cases of practical interest in which $r_0\gg \Delta \rho_0$, is much simpler than other nonparaxial extensions of the BG beams found in the literature, which rely on nonparaxial correction terms \cite{BG_1}, series of functions \cite{BG_2} or integral representations \cite{BG_3}.

\subsection{Circular Parabolic-Gaussian (CPG) beam}

\par If we apply the nonparaxial version of the envelope in Eq. \eqref{parabolic_gaussian_envelope} to Eq. \eqref{Azimuthally symmetric pattern as a function of A}, we get an azimuthally symmetric version of the PG beams, which we will refer to as Circular Parabolic-Gaussian (CPG) beams.

\par To illustrate, we will consider two examples. The first is an odd beam ($n=3/2$) with $\nu=h=1$ and $r_0=200\,\mu\text{m}$ while the second is an even beam ($n=1/2$) with $\nu=-3/2$, $h=0.5$ and $r_0=200\,\mu\text{m}$. Both are analyzed in two regimes: paraxial ($k_{\rho_0}=0.01k_0$) and nonparaxial ($k_{\rho_0}=0.5k_0$), with $\lambda=632.8\,\text{nm}$. 

\par
\begin{figure}[h!]
\centering
\includegraphics[width=0.65\columnwidth]{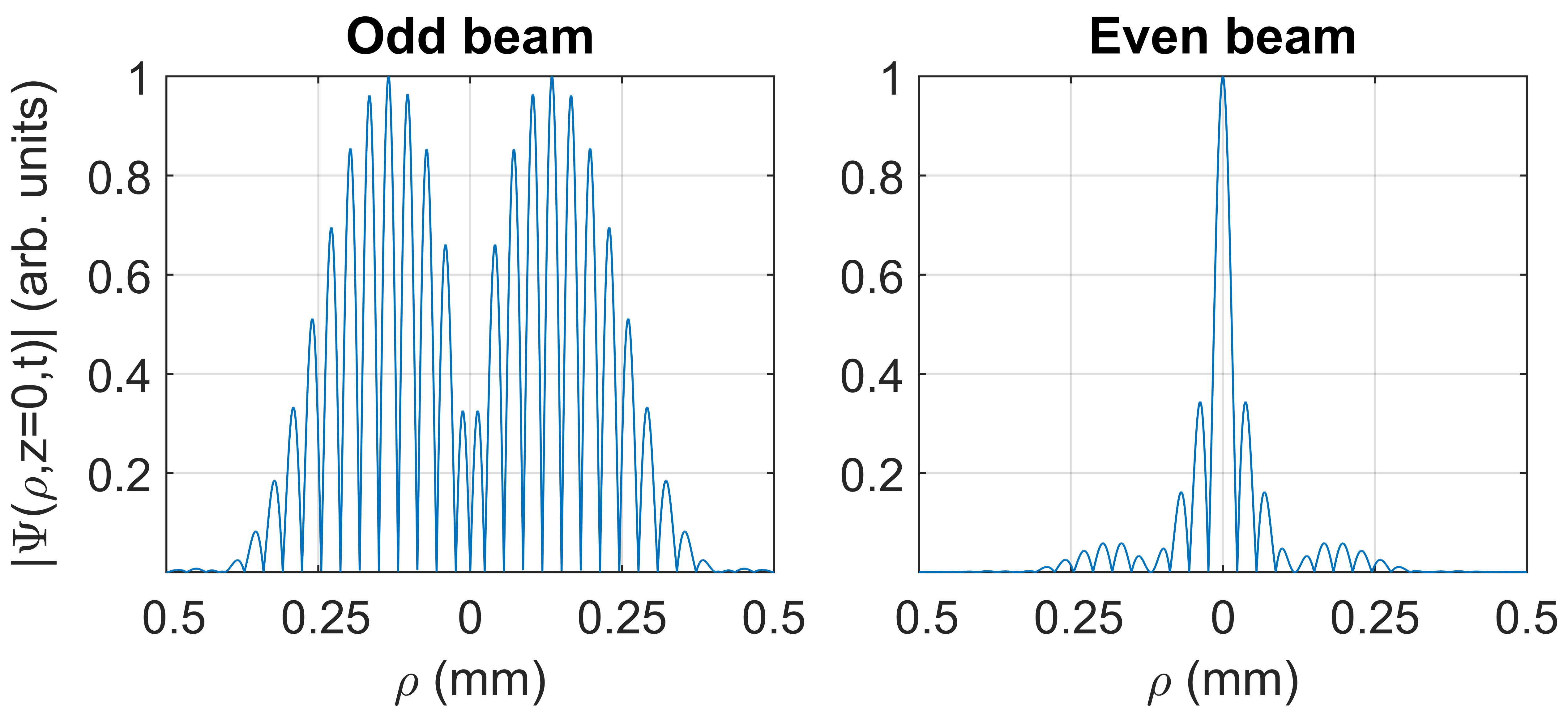}
\caption{Patterns of the CPG beams at $z=0$ for the paraxial case.}
\label{parabolic_initial}
\end{figure}

 Since the envelopes of PG beams have definite parity in $\rho$ at $z=0$, the initial patterns of CPG beams are Gaussian-apodized parabolic cylinder functions with an additional $1/\sqrt{2[\pi k_{\rho_0}\rho+e^{-(\pi-2)k_{\rho_0}\rho}]}$ decay factor modulated by a fast oscillating sinusoidal pattern with period $2\pi/k_{\rho_0}$, as explained in sec. \ref{similarities with cartesian case section}. Fig. \ref{parabolic_initial} shows the initial patterns of the CPG beams for the paraxial case, in which $\Delta \rho_0 \sim 16\,\mu\text{m}$. In the nonparaxial regime, the patterns are similar, but with much faster oscillations ($\Delta \rho_0 \sim 0.32\,\mu\text{m}$).

\par
\begin{figure}[h!]
\centering
\includegraphics[width=0.75\columnwidth]{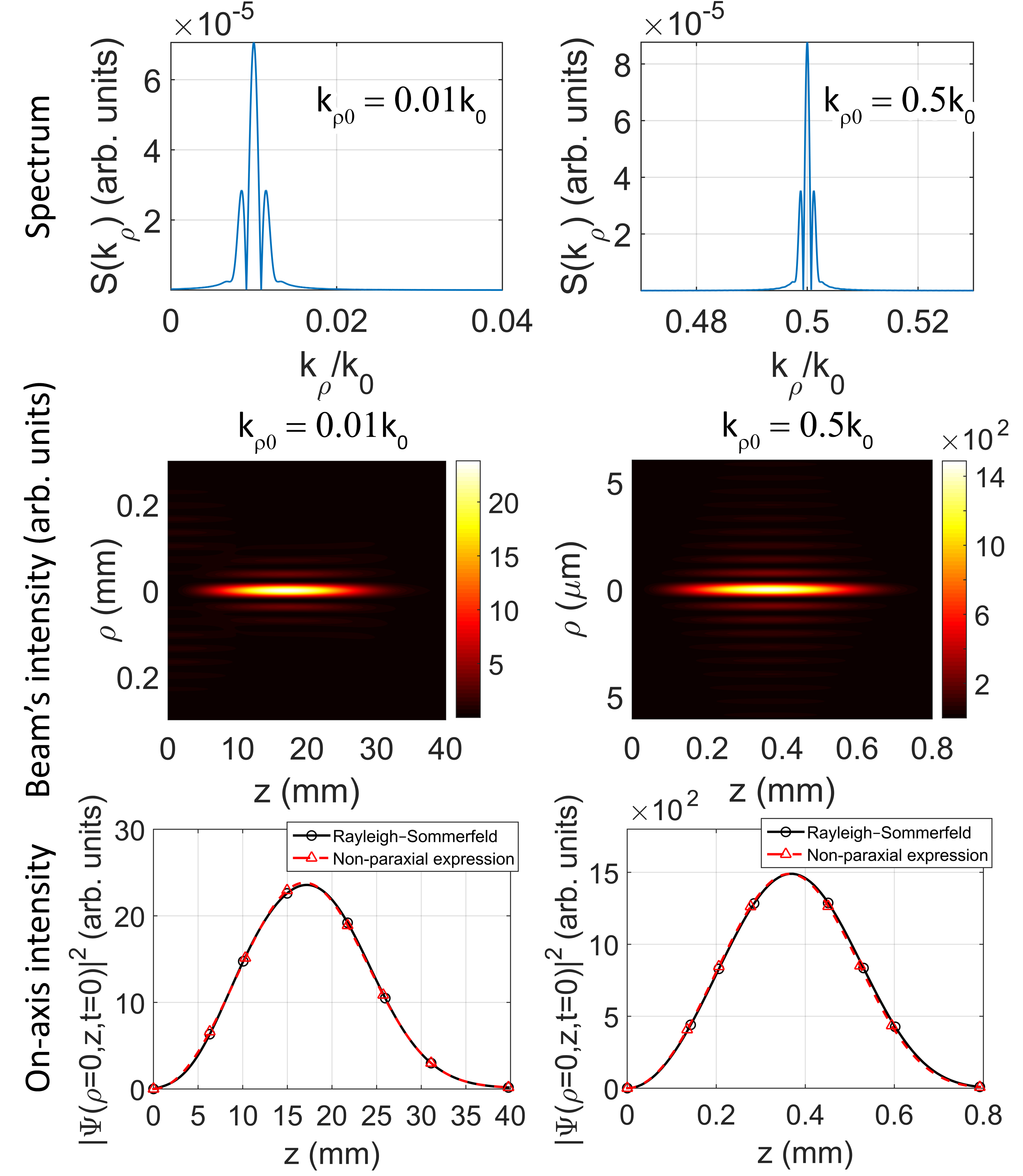}
\caption{Characteristics of a CPG beam with $n=3/2$, $\nu=h=1$ and $r_0=200\,\mu\text{m}$ in both paraxial ($k_{\rho_0}=0.01k_0$) and nonparaxial ($k_{\rho_0}=0.5k_0$) regimes for the wavelength $\lambda=632.8\,\text{nm}$. The first line shows the spectra, the second depicts the $|\Psi(\rho,z,t)|^2$ profiles according to the analytical expressions and the third compares the predicted on-axis intensities with the correct results calculated via Rayleigh-Sommerfeld diffraction integral.}
\label{parabolic_exemplo1}
\end{figure}

 Fig. \ref{parabolic_exemplo1} presents the resulting spectra and the intensity profiles of the odd beam. The first line depicts the spectra, calculated from the inverse Fourier transform of the initial profile (analogously to Eq. \eqref{S_A0}). Since the PG beams are solutions of the paraxial wave equation, their spectra are indeed concentrated and, in the case of their CPG counterparts, they are centered at $k_\rho=k_{\rho_0}$. The second line shows the resulting intensity profiles, normalized by their peak intensities at $z=0$. Although the odd symmetry makes the on-axis amplitude zero at $z=0$, the sidelobes shown in Fig. \ref{parabolic_initial} interfere constructively after some distance and create a very high-intensity peak over the axis, due to the concentration of the energy they carry in a small area. The effect is even more intense in the nonparaxial regime, in which the spot size is much smaller. The last line compares the on-axis intensity predicted by Eq. \eqref{on-axis-intensity-azymuthal} and the numerical calculation of Rayleigh-Sommerfeld diffraction integral with the initial pattern given by Eq. \eqref{Azimuthally symmetric wave at z=0}. It is clear that in both paraxial and nonparaxial regimes the analytical expression is accurate.

\pagebreak 

\pagebreak

\par
\begin{figure}[h!]
\centering
\includegraphics[width=0.75\columnwidth]{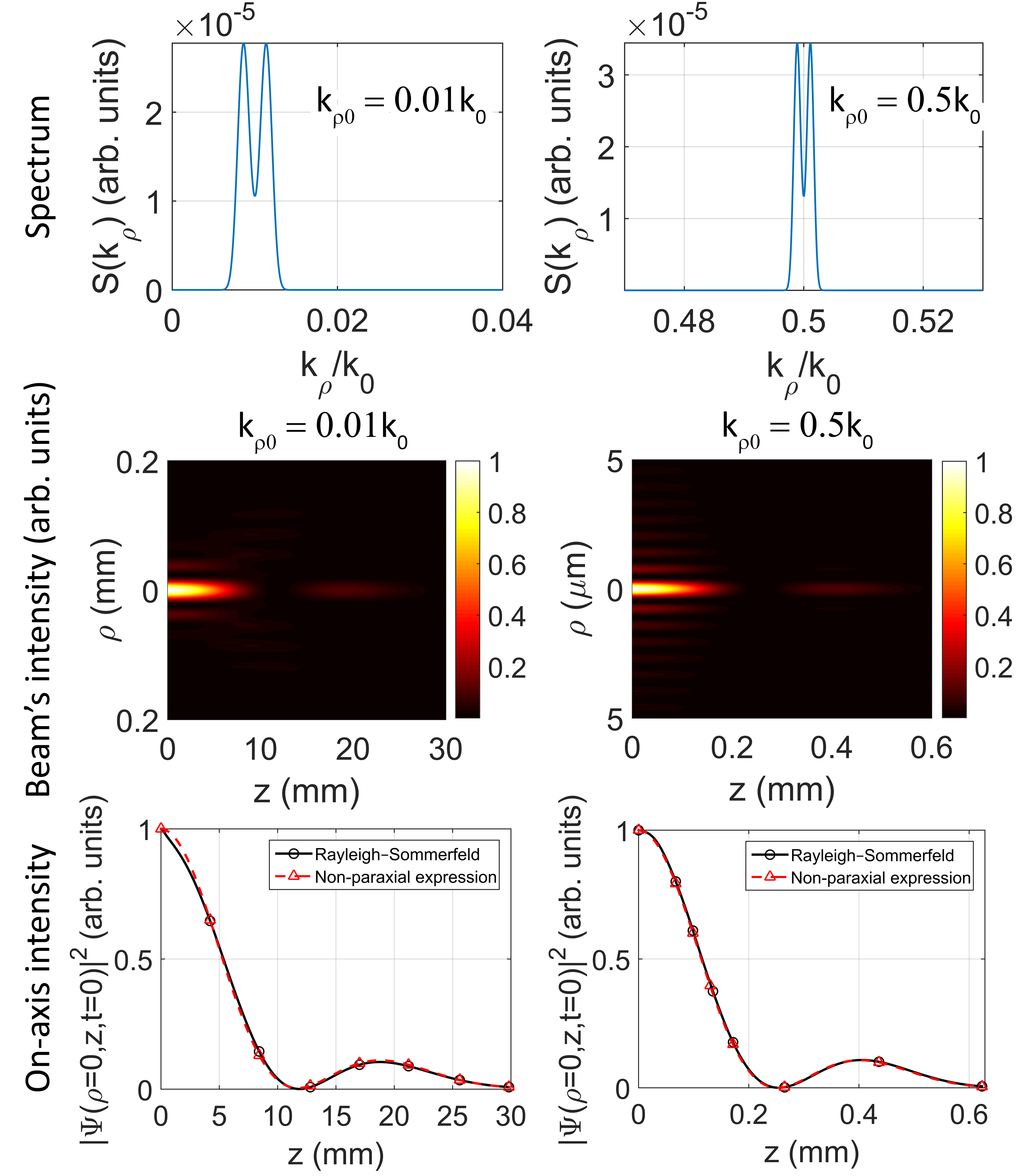}
\caption{Characteristics of a CPG beam with $n=1/2$, $\nu=-3.5$, $h=0.5$ and $r_0=200\,\mu\text{m}$ in both paraxial ($k_{\rho_0}=0.01k_0$) and nonparaxial ($k_{\rho_0}=0.5k_0$) regimes for the wavelength $\lambda=632.8\,\text{nm}$. The first line shows the spectra, the second depicts the $|\Psi(\rho,z,t)|^2$ profiles according to the analytical expressions and the third compares the predicted on-axis intensities with the correct results calculated via Rayleigh-Sommerfeld diffraction integral.}
\label{parabolic_exemplo2}
\end{figure}

 Fig. \ref{parabolic_exemplo2} presents the same kinds of results of Fig. \ref{parabolic_exemplo1} but for the even beam. We see again that the spectra are concentrated and that the on-axis intensities predicted by the analytical expressions are accurate. However, the even symmetry makes the peak intensities appear at $(\rho,z)=(0,0)$. Therefore, due to the normalization, we do not see higher-intensity on-axis peaks as in Fig. \ref{parabolic_exemplo1}.

\par The field depths of the examples presented can be estimated by Eq. \eqref{field_depth_expression} without recurring to the plots of intensity profiles. We could use $w_0=r_0$ for a reasonable estimate, but better results are obtained by choosing $w_0$ based on the envelopes of Fig. \ref{parabolic_initial}. For the odd beam, $w_0=0.35\,\text{mm}$ embraces all its significant lobes, while the same happens for the even beam for $w_0=0.3\,\text{mm}$. Using these values in Eq. \eqref{field_depth_expression}, we get $Z\approx 35\,\text{mm}$ and $Z\approx 0.56\,\text{mm}$ for the odd beam under paraxial and nonparaxial regimes, respectively, while $Z\approx 30\,\text{mm}$ and $Z\approx 0.48\,\text{mm}$ for the even beam under paraxial and nonparaxial regimes, respectively.

\section{Conclusions}

\newpage

\par In this work, we developed a theoretical analysis to efficiently describe the propagation of superpositions of waves with concentrated wavevector and frequency spectra, thus allowing a simple analytical description of fields with interesting transverse profiles. Starting with an extension of the paraxial formalism, we show how it can be applied to easily handle combinations of waves traveling in different directions without recurring to coordinate rotations. Moreover, and more importantly, we used a similar procedure to create a novel analytical description of azimuthally symmetric waves with concentrated spectra that can leverage all the previously presented results for unidimensional Cartesian waves, allowing us to build azimuthally symmetric versions of them and, therefore, to possibly derive new types of waves. Since these waves are composed of superpositions of zero-order Bessel beams with close cone angles that can be as large as desired, unlike in the paraxial formalism, we showed that it can also provide simple expressions for nonparaxial versions of known beams, such as the Bessel-Gauss beam. Finally, the validity and accuracy of the theory was corroborated by experimental demonstrations and comparisons to numerically-calculated Rayleigh-Sommerfeld diffraction integrals.

\section{Funding information}

S\~{a}o Paulo Research Foundation (FAPESP) (2015/26444-8); National Council for Scientific and Technological Development (CNPq) (304718/2016-5).

\section{Acknowledgments}

The authors thank Pedro Paulo Justino da Silva Arantes for his assistance with the numerical calculations of the diffraction integrals.

\bibliographystyle{abbrv}
\bibliography{mybib}

\end{document}